\documentclass[fleqn,10pt,twocolumn]{wlscirep}
\usepackage[utf8]{inputenc}
\usepackage[T1]{fontenc}
\usepackage{tikz}
\usetikzlibrary{tikzmark}
\usepackage{float}
\usepackage{amsmath,nccmath,amssymb}
\usepackage{braket}
\usepackage{extpfeil}
\usepackage{soul}
\usepackage [english]{babel}
\usepackage [autostyle, english = american]{csquotes}
\usepackage[switch]{lineno}
\MakeOuterQuote{"}%Makes quotation marks work correctly in terms of directions
\graphicspath{ {Figures/}}
\usepackage{gensymb}
\usepackage{miller}
\let\originallesssim\lesssim
\let\originalgtrsim\gtrsim

\let\vec\mathbf

\DeclareRobustCommand{\lesssim}{%
  \mathrel{\mathpalette\lowersim\originallesssim}%
}
\DeclareRobustCommand{\gtrsim}{%
  \mathrel{\mathpalette\lowersim\originalgtrsim}%
}

\makeatletter
\newcommand{\lowersim}[2]{%
  \sbox\z@{$#1<$}%
  \raisebox{-\dimexpr\height-\ht\z@}{$\m@th#1#2$}%
}
\makeatother

\begin{document}
\title{Investigation of the monopole magneto-chemical potential in spin ices using capacitive torque magnetometry}
\author[1,$\dagger$]{Naween Anand}
\author[1,2,$\ddagger$]{Kevin Barry}
\author[1,2,\S]{Jennifer N. Neu}
\author[1]{David E. Graf}
%\author[1]{Ju-Hyun Park}
\author[3]{Qing Huang}
\author[3]{Haidong Zhou}
\author[1,4]{Theo Siegrist}
\author[1,2]{Hitesh J. Changlani}
\author[1,2,*]{Christianne Beekman}
\affil[1]{National High Magnetic Field Laboratory, Tallahassee, FL 32310, USA}
\affil[2]{Florida State University, Department of Physics, Tallahassee, FL 32306, USA}
\affil[3]{University of Tennessee, Department of Physics, Knoxville, TN 37996, USA}
\affil[4]{Florida Agricultural and Mechanical University and Florida State University, College of Engineering, Tallahassee, FL 32310, USA}

\affil[$\dagger$]{present address: Intel Corp., Hillsboro, OR 97124, USA}
\affil[$\ddagger$]{present address: Ateios Systems, Newberry, IN 47449, USA}
\affil[$\S$]{present address: Oak Ridge National Laboratory, Nuclear Nonproliferation Division, Oak Ridge, TN 37831, USA}
\affil[*]{corresponding author: beekman@magnet.fsu.edu}

\makeatletter
\renewcommand{\@maketitle}{%
{%
\thispagestyle{empty}%
\vskip-36pt%
{\raggedright\sffamily\bfseries\fontsize{20}{25}\selectfont \@title\par}%
\vskip10pt
{\raggedright\sffamily\fontsize{12}{16}\selectfont  \@author\par}
\vskip25pt%
}%
}%
\makeatother
%\flushbottom
%
%\thispagestyle{empty}

\maketitle

\section*{Abstract}
The single-ion anisotropy and magnetic interactions in spin-ice systems give rise to unusual non-collinear spin textures, such as Pauling states and magnetic monopoles. The effective spin correlation strength ($J_{eff}$) determines the relative energies of the different spin-ice states. With this work, we display the capability of capacitive torque magnetometry in characterizing the magneto-chemical potential associated with monopole formation. We build a magnetic phase diagram of Ho$_2$Ti$_2$O$_7$, and show that the  magneto-chemical potential depends on the spin-sublattice ($\alpha$ or $\beta$), i.e., the Pauling state, involved in the transition. Monte-Carlo simulations using the dipolar-spin-ice Hamiltonian support our findings of a sublattice-dependent magneto-chemical potential, but the model underestimates the $J_{eff}$ for the $\beta$-sublattice. Additional simulations, including next-nearest neighbor interactions ($J_2$), show that long-range exchange terms in the Hamiltonian are needed to describe the measurements. This demonstrates that torque magnetometry provides a sensitive test for $J_{eff}$ and the spin-spin interactions that contribute to it. 

\section*{Introduction}

\begin{figure*}[!ht]
\centering
\includegraphics[width=0.6\textwidth]{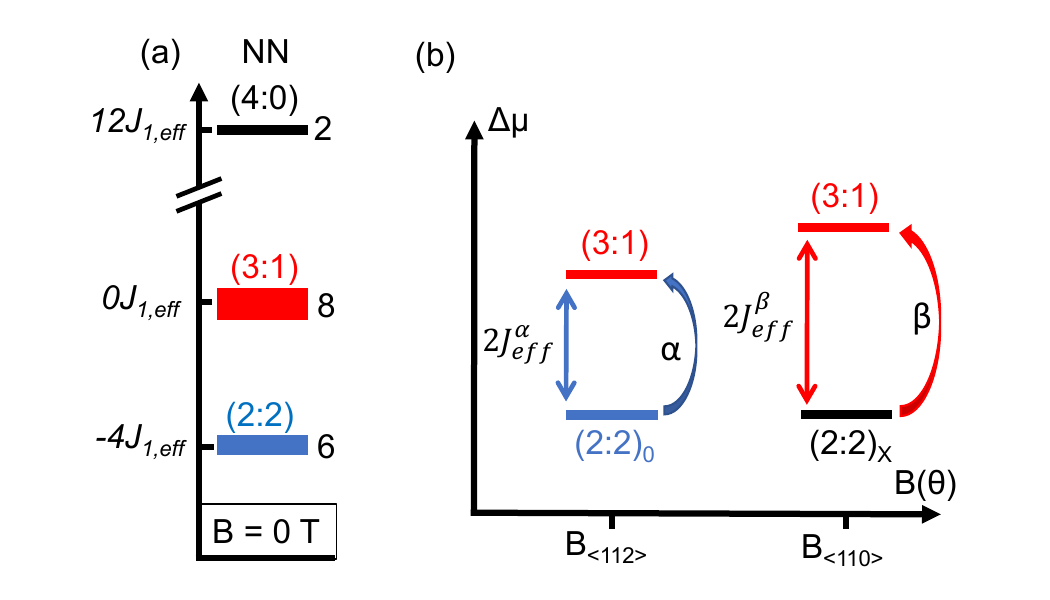}
\caption{{\bf Magneto-chemical potential (MCP) of spin-ice states.} (a) Schematic of the total energy of the pyrochlore lattice per tetrahedron (NN-model) for the spin-ice states at zero-field (we assign -$J_{1,eff}$ to an in-out pair and +$J_{1,eff}$ to an in-in/out-out pair) \cite{Bramwell_2020}. Note, if an isolated tetrahedron is considered, the energy of the (2:2) states would be -2$J_{1,eff}$. Each tetrahedron could adopt one of six possible (2:2) Pauling states at low temperatures. Monopole states, i.e., a (3:1) tetrahedron (eight possible configurations), reside at higher energy and these states freeze out at low temperature. The all-in/all-out (4:0) (2 configurations) are at much higher energy. (b) Schematic of the chemical potential difference ($\Delta\mu$) between the Pauling states ((2:2)$_0$ and (2:2)$_X$) and the (3:1) state. The MCP, 2$J_{eff}^\alpha$ and 2$J_{eff}^\beta$, depends on the field direction, i.e., whether monopoles nucleate on the $\alpha$- or $\beta$-spin sublattice (defined in Fig. \ref{spinconfig}) for $\vec{B}$||\hkl<112> and $\vec{B}$||\hkl<110>, respectively.  %An applied external field, depending upon its direction, allows HTO to attain several ordered phases. %The observed changes in the MCP of the (3:1) state as a function of field direction (with $B~\geq~2~T$) is summarized in panel (a). 
}
\label{Fig1:data}
\end{figure*}

 Geometrically frustrated systems have an inherent incompatibility between the lattice geometry and the magnetic interactions resulting in macroscopically degenerate ground-state manifolds.\cite{lacroix2011introduction,RogerGMelko,GardnerJ,Ramirez,Bramwell2001SpinMaterials} The large magnetocrystalline anisotropy and magnetic interactions in these systems give rise to unusual non-collinear spin textures, such as a spin-ice state that hosts emergent quasiparticle excitations equivalent to magnetic monopoles. \cite{CastelnovoMoessnerSondhi,Paulsen,Giblin,Fennell} As in ref.\cite{Bramwell_2020}, we denote the two-in/two-out Pauling states with (2:2), the 3-in/1-out monopole states as (3:1), and the all in/all-out configurations as (4:0). The effective spin-pair coupling ($J_{eff}$) determines the energy per tetrahedron for each of these states; for nearest neighbor exchange interactions, 2$J_{1,eff}$ is required to trigger the (2:2) $\rightarrow$ (3:1) transition. Importantly, the value of $J_{eff}$ is altered if interactions beyond nearest neighbor (i.e., dipolar \textit{D} and 2$^{nd}$ and 3$^{rd}$ nearest neighbor exchange \textit{J$_2$, J$_{3}$})  are included (see Table \ref{En-scales} and Fig. \ref{Fig1:data} (a)), as described in previously reported models \cite{Henelius,Ruff,Bramhartog,lacroix2011introduction,borzi}. Similar to applied biases controlling the electro-chemical potential of electrons in a material, an applied field %alters the MCP of the (3:1) states (i.e., the energy needed to add a monopole to a system in thermal equilibrium) can be altered by an applied field., thus making $J_{eff}$ field dependent (see Fig. \ref{Fig1:data}(a)). The applied field will 
 lowers the chemical potential of specific configurations leading to magnetic transitions between various non-collinear spin textures depending on the field direction and strength,\cite{Bramwell_2020,RogerGMelko,petrenko,FennellPetrenko,Harrisbramwell} (see Figs. \ref{Fig1:data} and \ref{spinconfig}).

Field-induced phase transitions in these systems have been studied by magnetometry, neutron scattering, ultrasound and dilatometry techniques, experimentally, or through numerical methods.\cite{Erfanifam,Saito,HARRIS1998757,ClancyRuff,RogerGMelko,Ruff, petrenko,Sato_2007,KreyLegl} In this work, we employ capacitive torque magnetometry (CTM) to characterize the spin-ice system Ho$_2$Ti$_2$O$_7$ (HTO) and to measure the effective spin-pair correlation strength between field-decoupled spins and the mean field. Conventional torque magnetometry is traditionally used to identify magnetic easy axes within crystalline materials.\cite{Phillips_1970} However, the large magneto-crystalline anisotropy\cite{Opherden_2019} makes HTO an ideal test-bed to reveal the unique capabilities of CTM in probing magnetic interaction energies, rather than the crystal field. From field dependent torque data, we extract the difference in MCP between the (2:2) and (3:1) states, i.e., the MCP of monopole creation. Note, in the field range of this study, these are transitions between ordered states, thus they cannot be classified as Kasteleyn transitions.\cite{Fennellpinchpoint,CastelnovoMoessner} 

A striking result is that the extracted MCP associated with monopole formation is different, depending on whether the monopoles nucleate on the $\alpha$- or $\beta$-spin sublattices\cite{ClancyRuff} (see Fig.\ref{Fig1:data}(b)). % In other words, the interaction strength between field-decoupled spins, $J_{eff}$, is field-direction dependent. 
While this conclusion is supported by the standard DSI-model-based classical Monte Carlo (MC) simulations (including dipolar interaction \textit{D}, see Table \ref{En-scales}), a comparison with the data clearly reveals the shortcomings of this form of the DSI Hamiltonian. Addition of a next-nearest neighbor exchange term ($J_2\sim$ 0.35~K) improves the correspondence between the simulated and measured torque data for the transition involving the $\alpha$-spins, providing an estimate for $J_2$. This term only marginally increases the stability (i.e., angular range) of the (2:2)$_X$ state, thus, a good agreement between the simulated and measured torque curves is still lacking for this transition. Additional third-nearest neighbor exchange terms, $J_{3}^a$ and $J_{3}^b$, are therefore required to fully describe the field-induced phase transitions in HTO. The idea of needing long-range exchange interactions for the complete description of spin ices is not new. Values for Dy$_2$Ti$_2$O$_7$ have been extracted via modeling of susceptibility and neutron data \cite{borzi, Samarakoon,Ruff,Henelius}, but to the best of the authors' knowledge, no such modeling has been reported for HTO.

\section*{Results}

{\bf Torque rotations in the \hkl(001) and \hkl(1-10) planes.} \\Torque magnetometry measurements have been performed on crystallographically oriented HTO single crystals as a function of external field strength, field direction, and temperature. Figs. \ref{Fig2-1:data}(a) and (b) show the torque responses when the field is rotated within the \hkl(001) and the \hkl(1-10) plane of the unit cell, respectively. The zero-field contribution has been subtracted for all curves to show the magnetic response of the system, which is characterized by multiple sharp turnovers and zero crossings, with intermediate sinusoidal responses that are directly related to different crystallographic axes of the spin-ice system (see Methods and the Supplementary Note 1).

%When rotating in the $x$-$y$ plane, the zero crossings are observed at intervals of 45\degree \ and sharp turnovers in the data occur near each of the \hkl<110> directions. The system always remains in one of the Pauling states in this plane, in particular $Q~=~0$, (2:2)$_0$, except near \hkl<110> directions when all $\beta$-spins flip and the net magnetization sharply rotates by 90\degree\ . At low field with the (3:1) states energetically out of reach, hysteresis appears around this transition, which is a clear sign of glassy behavior\cite{SM}. For the body diagonal plane, the torque response shows additional zero crossings and turnovers with significant changes in the observed torque amplitude when approaching different high symmetry axes. In the following, using a single unit cell model based on the previously observed spin textures, (Fig.\ref{Fig1:data}(b)-(d)), we unambiguously determine the angular and field stability of each of the magnetic phases hosted by the spin ice. From the angles at which (2:2) $\Leftrightarrow$ (3:1) transitions occur we determine the MCP (i.e., the Zeeman energy $E_Z$) associated with (3:1) state formation. 

\begin{figure*}[!ht]
\centering
\includegraphics[width=1\textwidth]{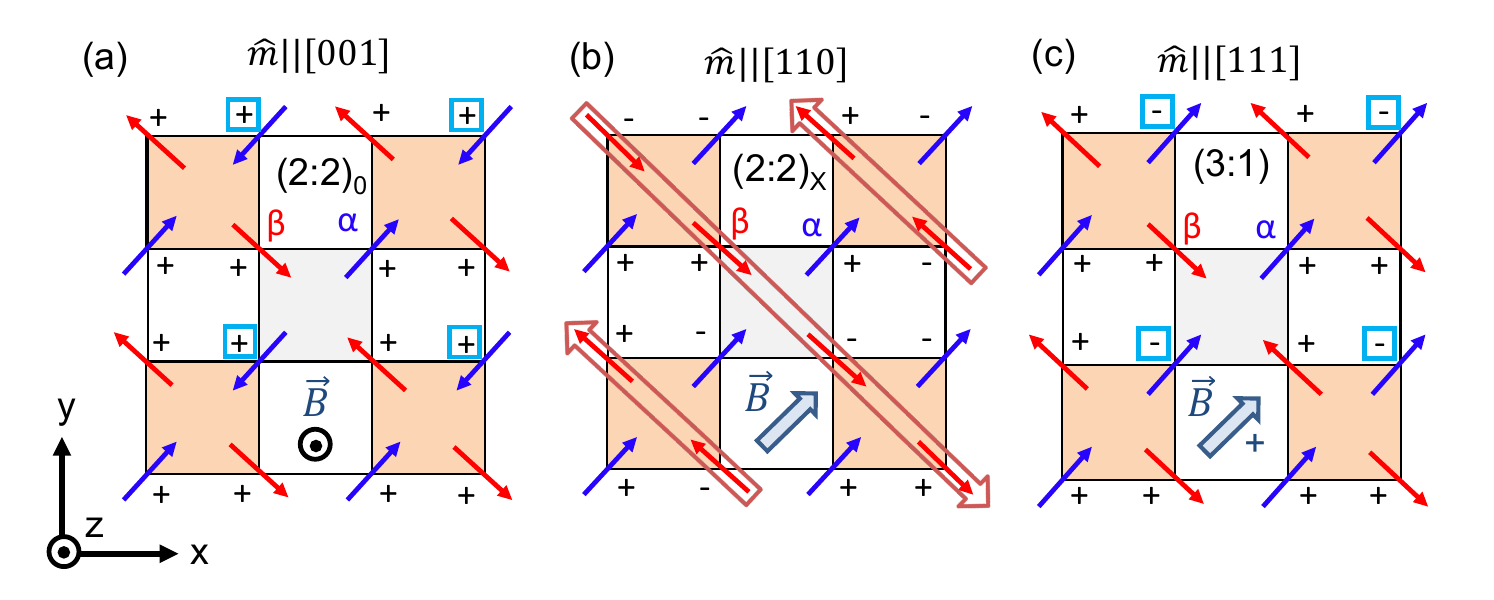}
\caption{{\bf Spin textures of the ordered spin-ice states for different applied fields}. Spin-ice magnetic phases displayed as a 2D projection of the HTO unit cell down the $z$-axis. The orange squares are tetrahedra. The square in the center (grey) is not a tetrahedron; the diagonally opposing spins are in the same lattice plane. The +/- signs indicate the spin directions along the $z$-axis. \cite{RogerGMelko,FennellPetrenko,Harrisbramwell,ClancyRuff} (a) (2:2)$_0$ state with a net magnetic moment ($\vec{\hat{m}}$) in the $z$-direction ((2:2)$_0$ states also form with $\vec{\hat{m}}$ in the $x$- or $y$-directions, depending on the field direction); (b) (2:2)$_X$ state (with $\vec{\hat{m}}$ $\parallel$\hkl[110]) in which the $\alpha$-spins (blue) are polarized and the $\beta$-spins (red) are antiferromagnetically aligned in chains (highlighted by the open arrows). The (2:2)$_X$ state also forms when $\vec{B}$ is directed along any of the family of \hkl<110> directions; and (c) (3:1) state with one spin flip per tetrahedron with $\vec{\hat{m}}$ $\parallel$\hkl[111] (the (3:1) state forms when $\vec{B}$ is directed along any of the family of \hkl<111> directions with |$B$|~$\geq$ 2 T). The spins denoted by the light blue boxes in panels (a) and (c) indicate the spin sublattice that becomes decoupled from the field when the magnetic field is directed exactly along the \hkl[112] direction. The direction of the magnetic field is indicated for each spin texture. The total energies per 16-site unit cell, i.e., interaction energies summed over 1$^{st}$, 2$^{nd}$ and 3$^{rd}$ nearest neighbors and the Zeeman energies, are calculated for each of these ordered state, details regarding this calculation are provided in Supplementary Note 7. }
\label{spinconfig}
\end{figure*}

\begin{figure*}[!ht]
\centering
\includegraphics[width=1\textwidth]{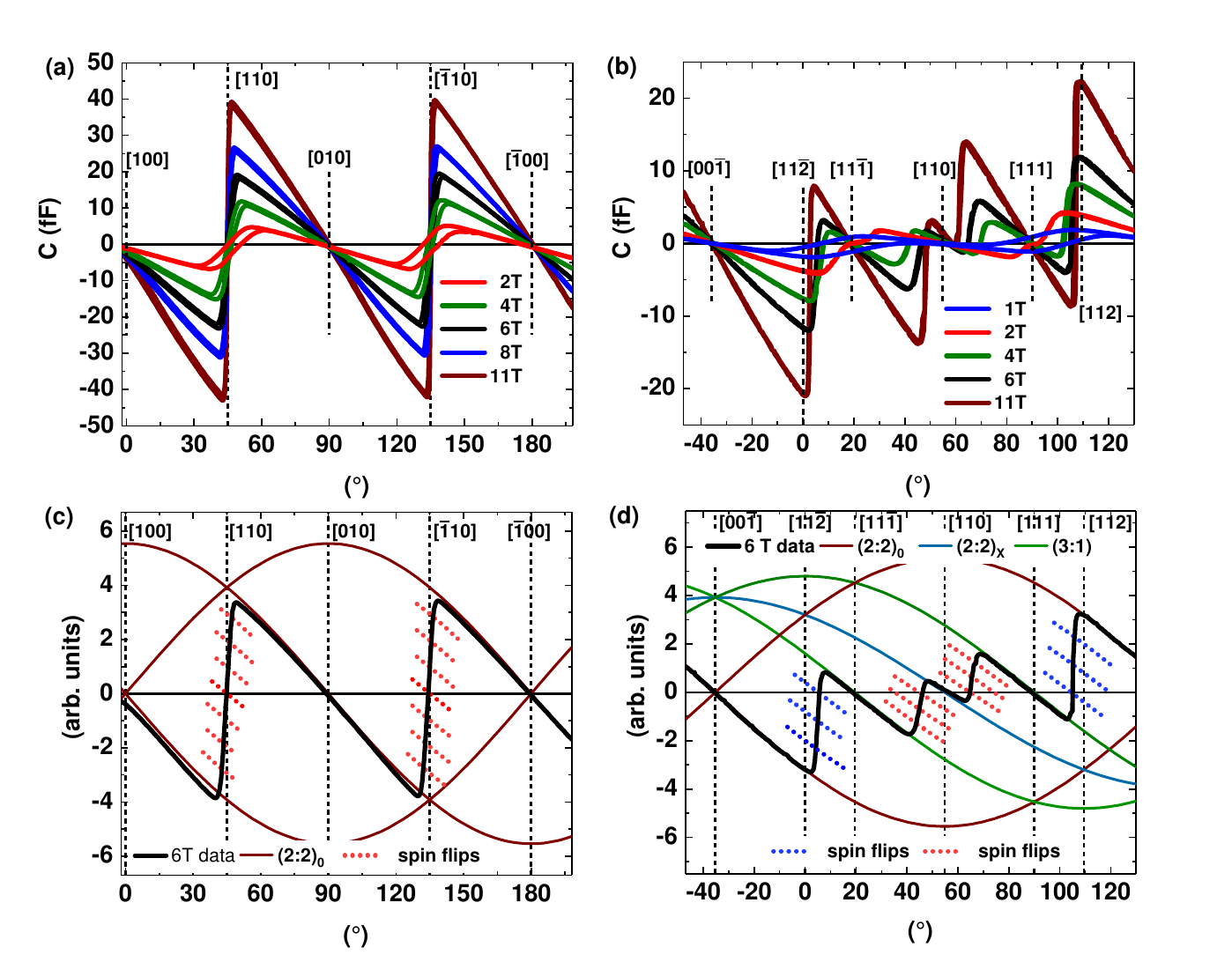}
\caption{{\bf CTM angular measurements.} Measured capacitance change as a function of angle in various applied fields for an HTO single crystal, measured at $T$ = 0.5 K. (a) CTM response when $\vec{B}$ rotates in the \hkl(001) plane (the \hkl[100] direction corresponds to 0\degree ). (b) CTM response when $\vec{B}$ rotates in the \hkl(1-10) plane (the \hkl[11-2] direction corresponds to 0\degree ). (c), (d) CTM response in a 6~T applied field for the \hkl(001) plane and the \hkl(1-10) plane, respectively. The 6-T data (solid black line) was scaled according to sample volume and cantilever sensitivity. The data is plotted alongside calculated torque curves using the phenomenological model described in the Methods section and the Supplementary Note 1 and Note 2. The solid colored model curves correspond to stable phases, while the dotted lines are calculated using volume fractions of spin flips on the $\alpha$ and $\beta$ sublattices. For each panel, crystallographic directions are indicated by vertical dashed lines.}
\label{Fig2-1:data}
\end{figure*}

A phenomenological single-unit-cell model is used to map out the evolution of magnetic phases as a function of field orientation for a given field strength. In this model, the sinusoidal torque curves are generated by explicitly calculating the torque response for one 16-site cubic unit cell for the different spin textures shown in Fig. \ref{spinconfig} (solid curves in Fig. \ref{Fig2-1:data} (c) and (d)) and intermediate mixed textures (dotted curves in Fig. \ref{Fig2-1:data} (c) and (d)). These curves provide an objective way to determine the half-way point of each of the transitions. Based on a comparison of the model curves and the data, the (2:2)$_0$ is the only stable phase in the \hkl(001) plane, except near the \hkl[110] and the \hkl[-110] directions when all $\beta$-spins flip and the net magnetization sharply rotates by 90\degree. At low field with the (3:1) states energetically out of reach, hysteresis appears around this transition, which is a clear sign of glassy behavior (see Fig. \ref{Fig2-1:data}(a) and the Supplementary Notes 1 and 5). 

For the rotation within the \hkl(1-10) plane all three spin textures show appreciable angular stability against misalignment of the field (see Fig. \ref{Fig2-1:data} (b) and (d)). While this may not be surprising for the (2:2)$_0$ and (3:1) phases, we find the (2:2)$_X$ phase to be strikingly stable around the [110] direction, especially in small applied fields. Although a single-unit-cell model is not adequate to describe the long-range antiferromagnetic alignment of the $\beta$-spins of this phase, its stability indicates that a long-range ordered phase is present around this crystallographic direction, rather than a transient domain state as observed in the \hkl(001) plane rotation (Fig. \ref{Fig2-1:data}(a) and (c)). [For field-angle phase diagrams, see Supplementary Figure 2.]

A way to visualize the surprising anisotropy in the (2:2)$_X$ stability between the two rotation planes, is to explore the energy surface that is obtained by integrating the torque curves. We show the energy surface contours associated with the \hkl(001) and the \hkl(1-10) rotation planes in the Supplementary Figure 3, with (2:2)$_X$ residing on a sharp maximum in the \hkl(001) plane and on a local minimum in the \hkl(1-10) plane. Thus, the (2:2)$_X$ phase resides on a saddle point in the energy landscape. While it is quite robust against misalignment of the field in the \hkl(1-10) plane, in the (001) plane the (2:2)$_X$ is not stable, and the system favors the (2:2)$_0$ states (i.e., a domain state with $\vec{\hat{m}}||\hkl[100]$ and $\vec{\hat{m}}||\hkl[010]$). The experimental observation of the (2:2)$_X$ phase is extremely sensitive to field misalignment, the high sensitivity of the CTM technique and the $<1\degree$ accuracy of the polished crystal faces proved critical for our measurements. 
%a 2D field-angle phase diagram is determined and shown in Fig.\ref{Fig2-1:data}(c) and (f). 
%In Fig.\ref{Fig2-1:data}(b) and (e), we compare the model calculated torque curves to the angular measurement in the $x$-$y$ plane and body diagonal plane taken at 6~T (black solid curves). The other solid-colored curves are associated with either the Pauling states or (3:1) spin configurations. The red or blue dashed curves in between are plotted to simulate the transitions between stable phases and are generated by assuming a changing volume fraction of spin flips on the beta or the alpha spin sub-lattices, respectively.\cite{SM} 

%As mentioned, in the \hkl(1-10) plane, there are phase transitions between the (2:2)$_0$ and (3:1) phases near the \hkl<112> directions, and between the (2:2)$_X$ and (3:1) states near the \hkl<110> directions.Starting with the first transition,
Within the \hkl(1-10) plane, the (2:2)$_0$~$\Leftrightarrow$~(3:1) transition occurs when the field rotates across the \hkl[112] (and \hkl[11-2]) direction, when $\vec{B}~||~\hkl[112]$ one $\alpha$-spin per tetrahedron (Fig.\ref{spinconfig}(a) and (c)) becomes decoupled from the applied field.\cite{Sato_2007,Matsuhira_2007,Ruff} The spins that are decoupled from the field maintain their spin-ice configuration due to the presence of the local internal field that is set by their spin environment. At a critical angle (i.e., a critical field) away from the \hkl[112] direction towards \hkl[111], the external field compensates the local internal field acting on the aforementioned spin-sublattice allowing them to flip. From these critical angles at which (2:2)~$\Leftrightarrow$~(3:1) transitions occur, we determine the MCP (i.e., the energy) associated with (3:1) state formation. Similarly, for the (2:2)$_X$~$\Leftrightarrow$~(3:1) transition, when the applied field is aligned along the \hkl[110] (or equivalent) direction, there exist two $\beta$-spins per tetrahedron, which are decoupled from the field. The unit cell still maintains the spin-ice configuration, however, that configuration is not unique, which leads to domains of degenerate magnetic phases. Theoretical and experimental evidence demonstrate the importance of second and third neighbor exchange couplings in addition to dipolar interactions \cite{borzi,Samarakoon, Ruff,Henelius}, but evidence linking these correlations to the antiferromagnetic alignment of the decoupled $\beta$-chains is still lacking. In other words, these beyond-NN exchange interactions that are reportedly needed to stabilize the predicted low temperature ordered phase \cite{borzi} involving alternating (single and double) spin chains, also play a role in stabilizing the (2:2)$_X$ phase at intermediate temperatures.  
We find ourselves well positioned to investigate the presence of these additional correlations because CTM allows us to extract the MCP of spin flip excitations for each of the sublattices separately. \\

\begin{figure*}[htp]
\centering
\includegraphics[width=0.87\textwidth]{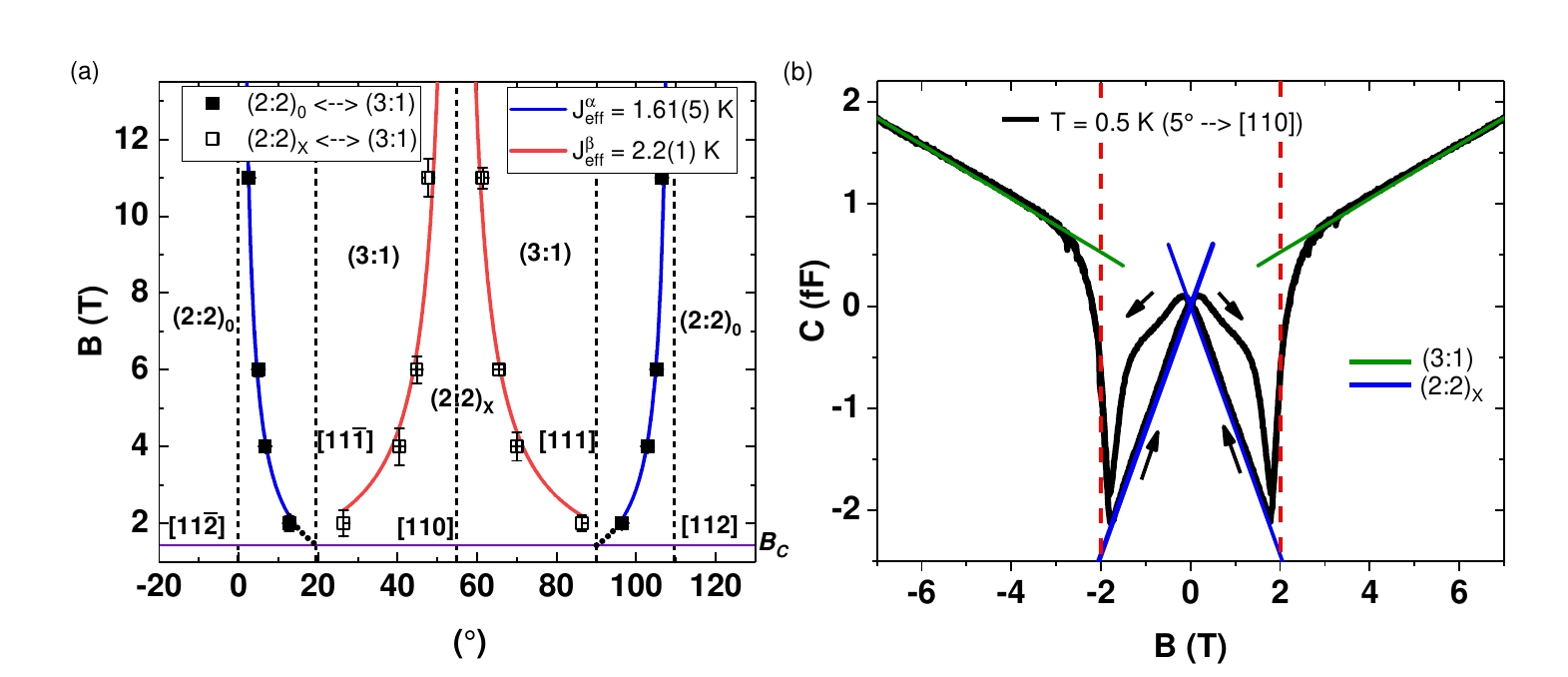}
\caption{{\bf Monopole MCP extraction from experimental torque data.} (a) Critical angles extracted from the experimental data in Fig. \ref{Fig2-1:data}, which define the transitions between the observed spin-ice magnetic phases, as a function of applied field (for the \hkl(1-10) plane rotations). The plotted error bars on the data points are a visualization of the residuals of the fits. Fits are shown for transitions between (2:2)$_{0}$~$\Leftrightarrow$~(3:1) (blue curves) as well as (2:2)$_{X}$~$\Leftrightarrow$~(3:1) (red curves). The errors on the extracted J$_{eff}$ values are uncertainties obtained from the fits. The black dotted lines represent the extrapolation of the fit curves to the nearby \hkl<111> directions. The horizontal purple line represents the location of the critical field required to transition between the Kagome ice and the (3:1) phases. Crystallographic directions are indicated by vertical dashed lines (b) Capacitance change as a function of applied field at $T$ = 0.5 K with $\vec{B}$ applied $\approx 5\degree$ away from the \hkl[111] direction, the black arrows indicate the sweep direction. The solid lines are volume-scaled calculated torque curves for the (3:1) (green) and the (2:2)$_{X}$ phase (blue) (see Supplementary Note 4 for more details). The dashed vertical lines indicate the critical field of the (2:2)$_{x}$ to (3:1) transition associated with a $J^{\beta}_{eff}$ = 2.1 K. }
\label{Fig3}
\end{figure*}

\noindent{\bf Monopole MCP extraction from CTM data.} \\The extracted critical angles are shown in the phase diagram in Fig. \ref{Fig3}(a) for both transitions. By identifying the field-decoupled spin sublattice for each transition, we fit the extracted angles as a function of applied field and extract the MCP associated with (3:1) monopole creation/annihilation. For the (2:2)$_0$~$\Leftrightarrow$~(3:1) phase transition, a value of $J^{\alpha}_{eff}$=1.61(5)~K is determined from the experiment (details on the analysis are provided in the Methods section). In addition, if one extrapolates the fitted curves to the nearby \hkl<111> directions, a crossing point occurs at $B_{c}$=1.44~T in each case. These crossing points match well with theoretical predictions\cite{Zvyagin,Isakov}, $B_{c}=6J^{\alpha}_{eff}/(g\mu_{B}\langle J_{z} \rangle)$ and with experimental results ($B_{m}\approx$ 1.5~T  \cite{KreyLegl}) of the critical field required for the Kagome ice $\rightarrow$ (3:1) phase transition, which occurs as a function of increasing field when $\vec{B}$ is perfectly aligned along any of the \hkl<111> directions. 

Strikingly, the same analysis for the (2:2)$_X$~$\Leftrightarrow$~(3:1) transitions, yields a larger value of $J^{\beta}_{eff}$=2.2(1)~K. We confirm this larger effective spin-pair coupling strength for the (2:2)$_0$~$\Leftrightarrow$~(3:1) phase transition via field sweep measurements, with the field purposefully misaligned away from the \hkl[111] direction (see Fig. \ref{Fig3}(b)). Surprisingly, the small misalignment of 5\degree \ away from the \hkl[111] direction (towards the \hkl[110] direction) stabilizes a low field (2:2)$_X$ phase (rather than Kagome ice, expected when the field is perfectly aligned with any of the \hkl<111> directions), which transitions into the high field (3:1) monopole phase above a critical field.\cite{Fennellpinchpoint} We extract a critical field of 2~T for this transition, i.e., $J^{\beta}_{eff}$=2.1~K, in line with the results from angular sweep torque data. While, the agreement between $J^{\alpha}_{eff}$=1.61(5)~K and the predicted $J^{s-DSI}_{1,eff}$ with long-range dipolar interactions (see Table \ref{En-scales} and ref. \cite{lacroix2011introduction}) is remarkable, the s-DSI model does not describe the (2:2)$_X$~$\Leftrightarrow$~(3:1) transitions very well. As we will show below, the inclusion of higher order exchange terms affects the phase boundary and the stability of the spin-ice phases associated with both transitions.
%Defining the transition as the moment where 50\% of the $\beta$-spins have flippe

Identical measurements were performed at $T$ = 1.7~K, above the spin-freezing temperature \cite{KreyLegl} (see the Supplementary Note 5). We find that beyond thermal smearing, the (2:2)$_X$ state is the only phase that changes significantly. This is evident from the change in slope of the torque curve around the \hkl[110] direction. This indicates deviation from a “clean” (2:2)$_X$ state due to thermal defects in the spin lattice at $T$~=~1.7~K, which further supports the conclusion that the stable phase observed in CTM around the \hkl[110] direction is indeed the (2:2)$_X$ state. \\

\noindent{\bf Monte Carlo simulated torque curves.}\\ Classical Monte Carlo (MC) simulations were performed for a pyrochlore cluster with $16 \times 4^3 = 1024$ spins and periodic boundary conditions. Simulated torque curves are compared to the experiments. The results for a strictly nearest neighbor model (NN, blue curve) and for the s-DSI (long-range dipolar) model (including Ewald summation, loop moves, and demagnetization effects \cite{RogerGMelko}, red curve) are shown in Fig. \ref{Fig4:MC}(a) (see Supplementary Note 6 for more details). While the data are well described within either model at high field, it is clear that the experimental observations at low field are not fully described by either of these models. In low fields, the s-DSI model does well in approximating the critical angle associated with the transitions, but it overestimates the angular stability of the (3:1) phase. In contrast, the NN model better approximates the (3:1) stability, but does less well with the critical angles. Most noticeable at higher fields, is that the stability of the (2:2)$_X$ phase is underestimated in both models.

We apply the same procedure for the extraction of the (3:1) MCP for each transition from the torque curves obtained from the MC simulations. The results for the s-DSI model (with $J_2$~=~0) are presented in Fig. \ref{Fig4:MC}(b). We obtain $J^{\alpha,MC}_{eff}$=1.4(2)~K and $J^{\beta,MC}_{eff}$=1.8(1)~K for the (2:2)$_0$ $\Leftrightarrow$ (3:1) and (2:2)$_X$ $\Leftrightarrow$ (3:1) transitions, respectively. The errors are based on the angular resolution (1\degree \ ) of the simulations. While the qualitative trend is correct, these $J_{eff}$ values differ from results shown in Fig. \ref{Fig3}(a). The value for the (2:2)$_0$ $\Leftrightarrow$ (3:1) transition is only slightly smaller than the $J^{\alpha}_{eff}$ extracted from the measurements. That said, we note that there is a spread in reported values for the NN exchange and the dipolar interactions for spin-ice systems in existing literature,\cite{Ruff, Samarakoon,Bramwell_2020, Bramwell2001SpinMaterials} which could cause such a discrepancy. Similar to our experimental findings, the simulated curves show that the (3:1) MCP is not the same, depending on the sublattice that the monopoles nucleate on during the transition. However, the $J^{\beta,MC}_{eff}$ extracted from the MC simulations for the (2:2)$_X$ $\Leftrightarrow$ (3:1) transition is significantly smaller (1.8~K, Fig. \ref{Fig4:MC}(b)) compared to our experimentally observed value of $J^{\beta}_{eff}$ = 2.2~K (see Fig. \ref{Fig3}(a)).  

In Fig. \ref{Fig4:MC}(c), a snapshot of the spin texture in a $2\times 2 \times 2$ unit cell structure is shown as a 2-dimensional projection projected down the $z$-axis, illustrating the spin texture as extracted from the MC simulation at $T$ =~0.5~K with $B$~=~4~T~||~\hkl[110] in the \hkl(1-10) plane. Under these conditions the ground state of the system is represented by a (2:2)$_X$ phase with no evidence of defects in the spin lattice. While the model does predict the correct ground state, it does not capture the entire extent of the angular stability of the (2:2)$_X$ phase. 

To extend the DSI model beyond just the nearest-neighbor and dipolar terms, the minimal way is to add a next nearest-neighbor $J_2$ interaction. The presence of $J_2$ does not change the energetics of the (3:1) phase, but for $J_2>0$ (see Methods) an additional Ising antiferromagnetic interaction is introduced. We have simulated curves for various J$_2$ values up to 0.04 meV ($\sim$ 0.464~K). In Fig. \ref{Fig4:MC} we plot the torque curve associated with the s-DSI model with $J_2\sim$0.35 K added to it. This term improves the agreement between the data and the MC simulations for the transition involving the $\alpha$-spins, now accurately approximating the (3:1) stability at low fields, providing an estimate for the size of $J_2$ for HTO. The value of $|J_2/J_1|$ found in this work is similar to (but higher than) the reported value for the sister compound Dy$_2$Ti$_2$O$_7$~\cite{Henelius, Samarakoon}. However, while the angular stability of the (2:2)$_X$ phase did appear to marginally increase, the quantitative value of the angular extent (see inset) is not explained by adding the $J_2$ term, indicating that interactions such as $J_3$, are necessary for a precise characterization of the Hamiltonian.

We support our findings with a short-range phenomenological model, which we use to evaluate the interaction energy for each spin-ice phase (see Supplementary Note 7 for more details). From this analysis, one can see what effect each of the interaction terms in the Hamiltonian has on the phase boundary of the field-induced magnetic phase transitions in HTO. In short, for the (2:2)$_0 \Leftrightarrow$ (3:1) transitions, the introduction of a $J_2$-term affects the interaction energy of the (2:2)$_0$ state, but does not impact the energetics of the (3:1) state. Effectively, $J_2$ partially negates the effects of long-range dipolar interactions. Note, adding $J_3$-terms affects both the (2:2)$_0$ and (3:1) states in the same way, thus this effect cancels out when evaluating the location of the phase boundary associated with this transition. (These $J_3$ terms correspond to two different kinds of third nearest neighbors, their couplings are referred to as $J_{3a}$ and $J_{3b}$, see Supplementary Note 7). For the (2:2)$_X \Leftrightarrow$ (3:1) transitions, the introduction of $J_2$ also does not affect the energetics of the (2:2)$_X$ phase, as the interaction energy associated with this term sums to zero (i.e., similar to the (3:1) phase). 
%Hence, it is obvious that the phase boundaries of the (2:2)$_X$ and (3:1) transitions are unaffected by the $J_2$ term. 
Hence, the phase boundaries of the (2:2)$_X$ $\Leftrightarrow$ (3:1) transitions are unaffected by the $J_2$ term, a finding broadly consistent with the MC simulations. 
However, the $J_3$ terms affect the (2:2)$_X$ and (3:1) phases differently, and are therefore important in determining the location of the phase boundary for this transition. 
%These $J_3$ terms correspond to two different kinds of third nearest neighbors (their couplings are referred to as $J_{3a}$ and $J_{3b}$, see Supplemental Note 7). 
Thus, this simple short-range model allows us to constrain the value for $J_{3a} + J_{3b}$ to a ball-park value of $\sim$ -0.014 meV (-0.16~K). %These $J_3$ terms correspond to two different kinds of third nearest neighbors (their couplings are referred to as $J_{3a}$ and $J_{3b}$, see Supplemental Note 7). 
The sign and the order of magnitude for $J_{3a} + J_{3b}$ are consistent with previously reported values for DTO \cite{Henelius}. 
%This result is also consistent with our MC simulations.} 

While this work provides estimates for the interaction terms for HTO, owing to the strongly correlated nature of the system, a full re-optimization of all exchange parameters may be needed. An accurate determination of the individual values for  $J_{3a}$ and $J_{3b}$ requires further extensive MC simulations, which we leave to future work. %We envisage that the sensitivity of the CTM measurements, possibly in combination with other methods such as neutron scattering, specific heat, and bulk magnetization will be instrumental in determining the optimal parameters for HTO. 

In conclusion, we have shown that CTM can be used to evaluate the phase boundaries of magnetic phase transitions in spin-ice systems. The unique nature of the pyrochlore lattice and the spin-ice interactions allows us to evaluate the effects of $J_2$ and $J_3$ terms of the Hamiltonian separately, i.e., by investigating different phase transitions. %Our work shows that in highly anisotropic systems torque magnetometry can be used to extract detailed magnetic phase diagrams and to obtain information about the local energy scales (i.e., magnetic interaction strengths) associated with magnetic phase transitions between spin textures. 
We believe that CTM may serve as a natural complement to neutron scattering, specific heat and magnetization measurements which can be compared with careful numerics~\cite{Zhang_Changlani,Bhardwaj}, as it can put stringent bounds on effective Hamiltonians and theories of magnetic materials, thereby aiding to complete the understanding of their low-energy properties and response to magnetic fields.

%Error analysis MC figure: point spacing in MC curves is 1 deg, I have taken 0.25deg as the error in the angle, I did not go with 0.5 because that would overestimate the error (1 deg is the x-spacing between the torque maxima for that transition). This error bar on the angle falls well within the symbol size used in panel b). J was the best fit value averaged over both transitions of the same kind. The error bar was propagated. The error on an individual J value was determined by fitting the critical angle curve based on the extremes determined by the error bar on the angle. So I got a max and min value of J, the spread in J gave the error bar I propagated to find the error of the average J value posted in the figure. Updated fig panels uploaded as pdfs.    }

\section*{Methods}
\subsection*{Single Crystal Growth}
Single crystal samples of HTO were grown using the optical floating-zone method. Ho$_{2}$O$_{3}$ and TiO$_{2}$ powders were mixed in a stoichiometric ratio and then annealed in air at 1450$^{\circ}$C for 40~h before growth in an optical zone furnace. The growth was achieved by zone melting with a pulling speed of 6~mm/h under 5~atm oxygen pressure. Single crystal x-ray diffraction experiments, taken on an Oxford-Diffraction Xcalibur-2 CCD diffractometer equipped with a graphite-monochromated MoK$_{\alpha}$ source, confirm the symmetry (Fd-3m) and lattice parameter of 10.0839(1) \AA~ at 293 K, consistent with previous reports \cite{GardnerJ} (see Supplementary Note 8 for more details).

Crystallographic orientation and specific axis alignment was performed using an Enraf Nonius CAD4 4-circle single crystal x-ray diffractometer equipped with graphite monochromated $\mathrm{Mo}_{k_{\alpha}}$ radiation. Single crystals used for torque magnetometry measurements were prepared as cubes with 1 mm edge length. Crystallographic axis alignment to within 1\degree of the vector normal for each of the 6 polished faces was then confirmed, using single crystal x-ray diffraction, as a final check for each sample.%, with the polished faces being within 1$\degree$ of well defined crystallographic directions. 

\subsection*{Capacitive torque magnetometry}
Capacitive torque magnetometry measurements were performed at the National High Magnetic Field Laboratory in an 18 T vertical-bore superconducting magnet with a 3He insert allowing for an operating temperature range between 250~mK and 70~K. A calibrated Cernox resistance temperature sensor was used throughout our measurements to determine the sample temperature. Each single crystal sample was mounted onto a flexible BeCu cantilever, constituting the top plate of the parallel plate capacitor in our setup, and placed in an externally applied magnetic field while at low temperature (a schematic of the torque setup is provided in the Supplementary Figure 1). The applied magnetic field induces a torque, \boldmath$\vec{\tau} = \vec{m} \times \vec{B}$, on the magnetic sample causing the cantilever to deflect. This deflection yields a change in measured capacitance $\Delta \mathrm{C} = \mathrm{C} - \mathrm{C}\textsubscript{0}$ that is collected experimentally, where $\mathrm{C}\textsubscript{0}$ is the capacitance value collected in zero applied magnetic field. Here the magnitude of the induced torque $|\vec{\tau}|$ is proportional to the change in capacitance ($|\vec{\tau}| \propto \ \Delta \mathrm{C}$) with a proportionality constant that is dictated by the elastic properties of the BeCu cantilever. An Andeen-Harling AH2700A Capacitance Bridge operating at frequencies between 1,000 and 7,000~Hz was used to collect the capacitance data during each measurement. The measurement probe used allowed for rotation of the sample over a range of $\sim$200\degree~ and a Hall Sensor was used to calibrate the sample rotation with respect to the applied magnetic field. Schematics of the \hkl(001) and \hkl(1-10) planes and the high symmetry axes that lie on these planes are provided in the Supplementary Figure 1.

\subsection*{Phenomenological Model}
In this study, we have employed a simple unit cell model to calculate the expected torque response as a function of angle for each of the stable spin textures (see Fig. \ref{spinconfig}) (2:2)$_0$, (2:2)$_X$ and (3:1), and for the intermediate phases hosting an appropriate volume fraction of these spin textures. During each of the phase transitions, as the field is rotated within the \hkl(1-10) plane, the field-decoupled spins will flip to form intermediate domain states eventually leading to a (3:1) monopole phase on all tetrahedra. Depending on the transition, these are either only $\alpha$ or only $\beta$ spins. %Note, that for the field rotation within the \hkl(001) plane, the (3:1) states are likely energetically out of reach and monopoles are never formed in the transition between different (2:2)$_0$ states leading to the observed hysteresis (see Fig. \ref{Fig2-1:data}(a) and the Supplemental materials\cite{SM}. 
When rotating within the \hkl(001) plane, the measured magnetic torque component is given as,

\begin{equation}\label{eqnS1}
\tau_{n} = \hat{n}\cdot\vec{\tau} = [0, 0, \bar{1}]\cdot(\vec{m}\times\vec{B}); \;\;\;\;\;  \vec{B} = B[\cos \theta, \sin \theta, 0]
\end{equation}

When rotating within the \hkl(1-10) plane, these torque curves were calculated in the following way,

\begin{subequations}
\begin{equation} \tau_{n} = \hat{n}\cdot\vec{\tau} = \frac{[\bar{1}, 1, 0]}{\sqrt{2}} \cdot(\vec{m}\times\vec{B}); \;\;\;\;\;\; \vec{B} = B[\cos \theta~\hat{j}^{'} + \sin \theta~\hat{k}^{'}]
\end{equation}
\begin{equation} 
\hat{i}^{'}=\frac{[1, \bar{1}, 0]}{\sqrt{2}} ; \;\;\; \hat{j}^{'}=\frac{[1, 1, \bar{2}]}{\sqrt{6}}; \;\;\; \hat{k}^{'}=\frac{[1, 1, 1]}{\sqrt{3}}\;\;\;\;\;\;\;\;\; (\hat{i}^{'}\times \hat{j}^{'}=\hat{k}^{'})
\end{equation}%\begin{equation}
%\end{equation}
\end{subequations}
Supplementary Tables 1 and 2 list all the moment vectors and the functional forms of the angular dependence of the torque curves for both rotational planes. 

\subsection*{Determination of $J^{\alpha}_{eff}$ and $J^{\beta}_{eff}$}
We examined the critical angles associated with each of the phase transitions observed in our torque vs. angle measurements. We define the critical angle to be marked by the location where half of all tetrahedra have a (3:1) configuration. This angular position is extracted by finding the crossing point between the data and the associated model curve. Next, we identify which specific spin sublattice(s) decouple from the field and would be expected to flip when transitioning between the phases (see Figs. \ref{Fig1:data} and \ref{spinconfig}). While one may expect that the $\alpha$- and $\beta$-spin sublattices decouple exactly at \hkl<112> and \hkl<110> field directions, respectively, the internal field produced by the mean field will shift that transition to a critical angle away from these crystallographic directions. Thus, the Zeeman energy ($E_{Z}$) associated with this critical angle is a direct measure of this internal field. We calculate the analytic form of the Zeeman energy of the field-decoupled spins as the field rotates across the \hkl[112] and \hkl[110] crystallographic directions, respectively. Expressing E$_Z$ in terms of applied field ($B$) and field direction ($\theta$), and realizing that E$_Z$~=~2$J_{eff}$, allows us to determine a fitting function for the field vs. critical angle data from which the change in MCP ($J^{\alpha}_{eff}$ and $J^{\beta}_{eff}$) associated with the proliferation of (3:1) tetrahedra can be determined. The results of the fitting are presented in Fig. \ref{Fig3} (a) as the blue and red curves. For each type of spin sublattice, the MCP ($J^{\alpha}_{eff}$ and $J^{\beta}_{eff}$) takes the form:

\begin{ceqn}
\begin{align}\label{Zeeman}
E_{Z} =\Delta\mu= -\vec{m} \ \cdot \ \vec{B} = -10 \mu_{B} \hat{S} \ \cdot \vec{B}
\end{align}
\end{ceqn}

\noindent where $\hat{S}$ represents the unit vector associated with the given spin sublattice of interest for the transition. This procedure allows the derivation of a functional form for $B(\theta)$, which is used to fit the extracted values for the critical angles as a function of external applied fields (see Fig. \ref{Fig3}(a)). For the transition between the (2:2)$_0$ and (3:1) phase near the \hkl[112] direction, 

\begin{ceqn}
\begin{align}
\Delta\mu = 2J^{\alpha}_{eff} = \frac{10\mu_{B}B}{\sqrt{18}} \ [4\cos(\theta) + \sqrt{2}\sin(\theta)]
\end{align}
\end{ceqn}
The fitting function describing the transition near the symmetry-related \hkl[11-2] direction, can be derived in the same way. 
For the transition between the (2:2)$_X$ and (3:1) phase near the [110] direction,
\begin{ceqn}
\begin{align}
\Delta\mu = 2J^{\beta}_{eff} = \frac{10\mu_{B}B}{\sqrt{18}} \ [-2\cos(\theta) + \sqrt{2}\sin(\theta)]
\end{align}
\end{ceqn}

For the field sweep torque measurement in Fig. \ref{Fig3}(b), the applied field was misaligned by $\sim$ 5\degree\ away from the \hkl[111] direction (towards the \hkl[110] direction), which stabilizes a low-field (2:2)$_X$ phase (rather than a Kagome ice, which is formed when the field is perfectly aligned with the \hkl<111> directions). The field sweep shows two markedly linear regimes when the field is swept from high field to zero. These linear regimes correspond to constant saturated magnetization values, the ratio between the slopes describing these linear regimes (high field: green curve; low field: blue curve) are in great agreement with the ratio of saturated magnetization expected for the (3:1) (5 $\mu_B$/Ho) and (2:2)$_X$ (4.1 $\mu_B$/Ho) phase, respectively.  The field sweep also shows hysteresis around zero field, indicating a glassy response to a change in polarity of the applied field (i.e., the reversal of the $\alpha$ spins). %Using our model, we calculate the expected linear torque curves after each $\alpha$ (blue dotted lines) and $\beta$ (red dotted lines) spin flips. 

\subsection*{Monte Carlo Simulations}%\vspace{-0.2in}
We have simulated torque responses using the generalized DSI model whose Hamiltonian is given by,
\begin{equation}
\begin{aligned}
%	H = {} & -J_{1} \sum_{\langle i,j \rangle} \tilde{\textbf{S}}_{i} \cdot \tilde{\textbf{S}}_{j} + D r^{3}_{nn} \sum_{i>j} \Big(\frac{\tilde{\textbf{S}}_{i} \cdot \tilde{\textbf{S}}_{j}}{|\textbf{r}_{ij}|^3} - \frac{ 3 (\tilde{\textbf{S}}_{i} \cdot \textbf{r}_{ij}) (\tilde{\textbf{S}}_{j} \cdot \textbf{r}_{ij})}{|\textbf{r}_{ij}|^5} \Big)\\
%	& \;\;\;\;\;\;\;\;\;\;\;\;\;\;\;\; - g \mu_B \sum_{i} \vec{B} \cdot \tilde{\textbf{S}_i} 
	H = -J_{1} \sum_{\langle i,j \rangle} \tilde{\textbf{S}}_{i} \cdot \tilde{\textbf{S}}_{j}  -J_{2} \sum_{\langle \langle i,j \rangle \rangle} \tilde{\textbf{S}}_{i} \cdot \tilde{\textbf{S}}_{j}  \\
  + D r^{3}_{nn} \sum_{i>j} \Big(\frac{\tilde{\textbf{S}}_{i} \cdot \tilde{\textbf{S}}_{j}}{|\textbf{r}_{ij}|^3} - \frac{ 3 (\tilde{\textbf{S}}_{i} \cdot \textbf{r}_{ij}) (\tilde{\textbf{S}}_{j} \cdot \textbf{r}_{ij})}{|\textbf{r}_{ij}|^5} \Big)
 - g \mu_B \sum_{i} \vec{B} \cdot \tilde{\textbf{S}_i} 
\label{eq:Ham}
\end{aligned}
\end{equation}

where $\tilde{\textbf{S}}_{i}$ are classical spin vectors with $|\tilde{\textbf{S}}_i|=1$. The tilde is used to indicate that the spins are constrained to point along the local \hkl<111> axis of the tetrahedra they belong to. $\textbf{r}_{i}$ is the real-space location of site $i$, $\textbf{r}_{ij} \equiv \textbf{r}_{i} - \textbf{r}_{j}$, $\langle i,j \rangle$ ($\langle \langle i,j \rangle \rangle$) refers to nearest-neighbor (next nearest-neighbor) bonds, $r_{nn}$ is the nearest-neighbor bond distance, $J_1$ ($J_2$) is the nearest-neighbor (next-nearest neighbor) interaction strength, and $D$ is the strength of the long-range dipolar term. The $i>j$ notation guarantees each of pair of spins is only counted once.
$g \mu_B$ is the size of the magnetic moment and  $\vec{B}$ is the applied magnetic field.

Our calculations were performed for finite-size pyrochlore clusters (16 atoms per simple-cubic unit cell) with $N_{spins} = 16 \times 4^3 = 1024$ lattice sites, and with periodic boundary conditions. For the nearest neighbor model, we set $J_1$=+5.40 K and $D=0$.
%The length of each edge of the cubic unit cell is taken to be $l=1$. In terms of this length, the coordinates of all 16 sites of the unit cell are enumerated in the Supplemental Materials \cite{SM}. 
%Since In the case of long range interactions, it is known that in 3D with dipolar interactions($1/r^3$) the sum is conditionally convergent. 
To deal with long-range magnetic dipolar interactions, the Ewald summation technique was employed to convert the real space sum in the Hamiltonian into two rapidly convergent series, one in real space and the other in momentum space. This Hamiltonian was then simulated with the Metropolis Monte Carlo algorithm, using a combination of single spin flip and loop moves \cite{RogerGMelko} (which allows a ring of spins to flip at one time, while maintaining the ice rule constraint). For the s-DSI simulation (with long-range dipolar interaction $D$, red curve in Fig. \ref{Fig4:MC}(a)), the parameters were set to $J_1$ = -1.56~K, $D$ = 1.41~K and $g$=10 \cite{Bramwell2001SpinMaterials,Bramwell_2020}. $J_2$ was varied to investigate its effect on the angular extents of the (2:2)$_X$ phase. To produce the green curves in Fig. \ref{Fig4:MC}, $J_2$=0.35~K was used, additional simulations using different values for $J_2$ are presented in the Supplementary Figure 6. 
%Loop moves, allowing a ring of spins to flip at one time,  
Demagnetization effects (assuming a spherical sample) were taken into account \cite{RogerGMelko} for the presented simulated torque curves. 
More details of our simulations can be found in the Supplementary Note 6.

\section*{Data Availability}
The authors declare that the main data supporting the findings of this study are available within the paper and its Supplementary Information. The crystallographic data has been deposited with the joint CCDC/FIZ Karlsruhe online deposition service under nr. CSD-2172269.\cite{CSD} Other data that support the findings of this study are available from the corresponding author upon reasonable request.
%TC:ignore

\section*{Code Availability}
The code used to generate the Monte Carlo simulation results shown in the paper is publicly available at \url{https://github.com/hiteshjc/Ising_Ice_dipolar}
Additional scripts and files for the numerical calculations are available from H.J.C. upon reasonable request.

%\bibliography{TorqueHTO}

\section*{Acknowledgements}

C.B. acknowledges support from the National Research Foundation, under grant NSF DMR-1847887. J.N. and T.S. acknowledge support from the National Research Foundation, under grant NSF DMR-1606952. A portion of this work was performed at the National High Magnetic Field Laboratory, which is supported by National Science Foundation Cooperative Agreement No. DMR-1157490, No. DMR-1644779, and the State of Florida. H.D.Z acknowledges support from the NHMFL Visiting Scientist Program, which is supported by NSF Cooperative Agreement No. DMR-1157490 and the State of Florida. H.J.C. acknowledges support from the National Research Foundation, under grant NSF DMR-2046570, and start-up funds from Florida State University and the National High Magnetic Field Laboratory. The simulations were performed on the Research Computing Cluster (RCC) and the Planck cluster at Florida State University. We thank R. Moessner and L. Jaubert for helpful discussions.

\section*{Author contributions}

C.B. conceived the experiment(s) and analyzed the results, N.A. and K.B. contributed equally to this work, they conducted the torque magnetometry measurements and analyzed the results, D.G. assisted in conducting the torque magnetometry measurements, Q.H. and H.Z. synthesized the single crystals, J.N. and T.S. oriented and polished the crystals. H.J.C. performed the Monte Carlo simulations and contributed to the theoretical analysis. All authors reviewed the manuscript. 

\section*{Competing Interests}
The authors declare no competing interests
\clearpage
\onecolumn
\section*{Tables}
 \begin{table*}[htpb]
 \def\arraystretch{1.5}
\begin{center}
\caption{{\bf Interaction energy scales for HTO.} Values for the effective spin-pair interactions in HTO are provided for nearest neighbor model (NN), the standard dipolar spin-ice model (s-DSI, NN-dipolar \cite{Bramhartog}), and for the s-DSI model with long-range dipolar interactions (i.e., Ewald Summation) \cite{lacroix2011introduction,Ruff,RogerGMelko}. The generalized DSI (g-DSI) model includes couplings up to third neighbors, which have also been shown to depend on phonon-induced distortions to the lattice \cite{borzi}. In the g-DSI model, values have only been reported for Dy$_2$Ti$_2$O$_7$. The values presented in the bottom row are based on this work.}\label{En-scales}
\begin{tabular}{|c|c|}
\hline
 Model                   &   Energy Scales (K)\tabularnewline
\hline\hline
NN (no dipolar)   &  $J_1$= 5.40~~~~~    $J_{1,eff}$ = 1.8\\\hline 
Standard-DSI (NN-dipolar)    & $J_1^{s-DSI}$=  -1.56~~~~~     $D$ =1.41   ~~~~~  $J_{1,eff}^{s-DSI}$ = $\frac{J_1 + 5D}{3}$=1.83 \\\hline
Standard-DSI (long-range dipolar)      & $J_1^{s-DSI}$=  -1.56~~~~~     $D$ =1.41   ~~~~~  $J_{1,eff}^{s-DSI}$ = $\frac{J_1 + 4.53D}{3}$= 1.61 \\\hline
Generalized-DSI (long-range exchange)      & $J_{1}$=  -1.56~~ $D$=1.41~~  $J_{2}$ $\sim$0.35 ~~ $J_{2,eff}$= $\frac{J_2 - D/\sqrt{3}}{3}$=-0.155 ~~~~~ $J_3^a$  = ?~~~~~$J_3^b$  = ? \\\hline
\end{tabular}
\end{center}
\end{table*}

 \section*{Figures}
 
\begin{figure*}[h!]
\centering
\includegraphics[width=0.85\textwidth]{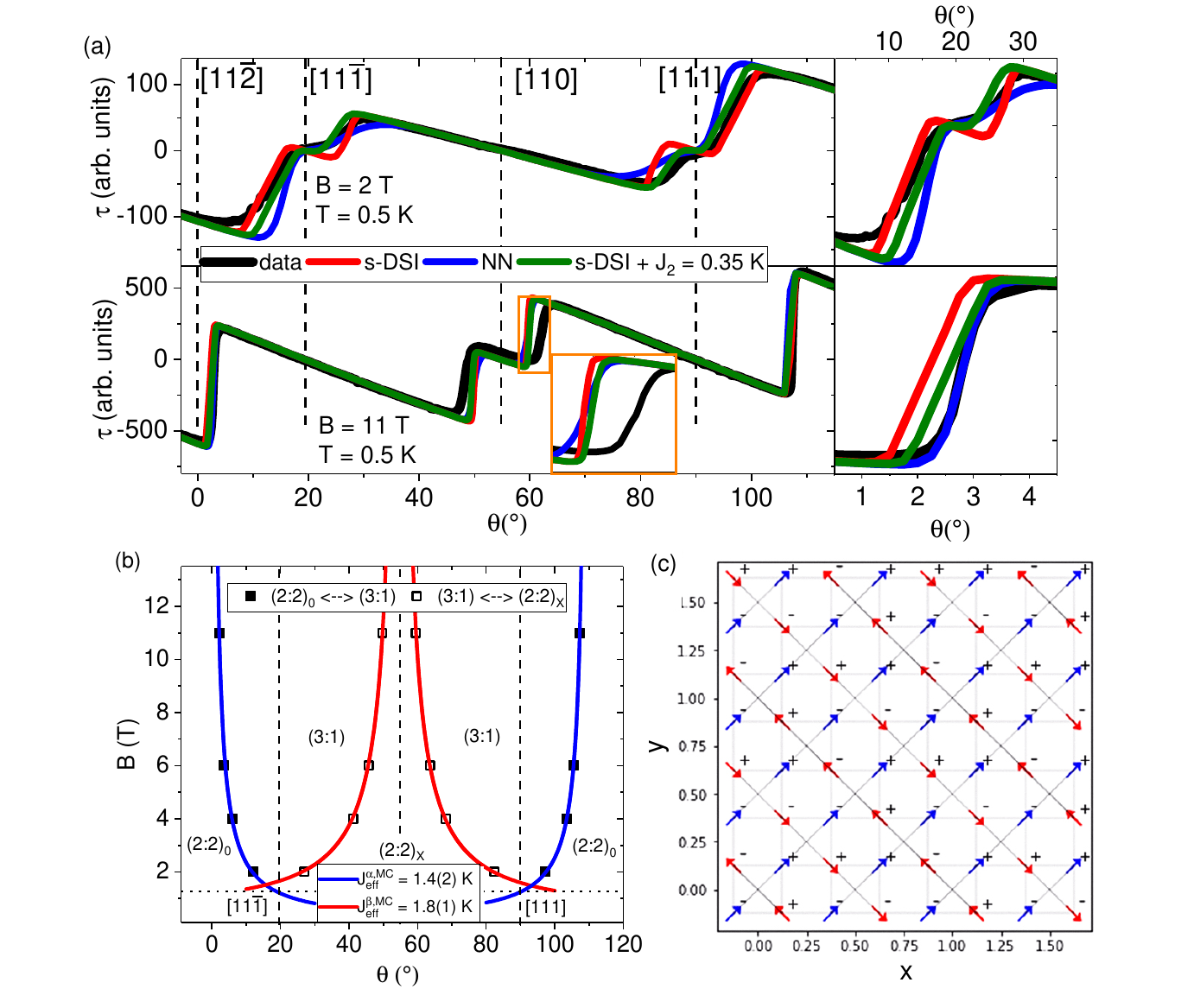}
\caption{{\bf Monopole MCP extraction from MC-simulated torque curves.} (a) Simulated torque response as a function of angle, obtained from MC simulations using the nearest neighbor model (NN, blue curves), s-DSI model (red curves) and the generalized DSI model (s-DSI with J$_2$ = 0.35 K, green curves); measured CTM data (black curves) are taken at $T$ = 0.5~K in applied fields of 2~T (top panel) and 11~T (bottom panel). The inset highlights the angular region near the \hkl[110] direction (orange square). The panels on the right highlight the region around the \hkl[11-2] direction. Crystallographic directions are indicated with vertical dashed lines. (b) Critical angles extracted from the torque curves obtained from the MC simulations (s-DSI, with long range dipolar $D$ and with $J_2$~=~0) for the transitions between the observed spin-ice magnetic phases, as a function of applied field (\hkl[1-10] plane rotations). Error bars are based on the angular resolution (1\degree \ ) of the simulations and are smaller than the symbol size.  Fits are shown for transitions between (2:2)$_0$ $\Leftrightarrow$ (3:1) (blue curves) as well as (2:2)$_X$ $\Leftrightarrow$ (3:1) (red curves). The uncertainties in the extracted $J^{\alpha,MC}_{eff}$ and $J^{\beta,MC}_{eff}$ values are errors determined from the fits. (c) A 2D snapshot of the spin texture (2x2x2 unit cells projected down the $z$-axis) taken during the MC simulation using $B$ = 4~T, at $\theta$ = 50\degree \ , and with $T$ = 0.5~K. The blue and red arrows indicate the $\alpha$- and $\beta$-spin sublattices, respectively.  The +/- signs indicate the spin directions along the $z$-axis. Under these conditions the ground state of the system is represented by a (2:2)$_X$ phase with no evidence of defects in the spin lattice.}
\label{Fig4:MC}
\end{figure*}
%\section*{Additional information}
%To include, in this order: \textbf{Accession codes} (where applicable); \textbf{Competing interests} (mandatory statement). 
%The corresponding author is responsible for submitting a \href{http://www.nature.com/srep/policies/index.html#competing}{competing interests statement} on behalf of all authors of the paper. This statement must be included in the submitted article file.
%TC:endignore
\end{document}

\maketitle

\section{The \hkl(001) rotational plane}

\begin{figure*}[h!]
\centering
\includegraphics[width=\textwidth]{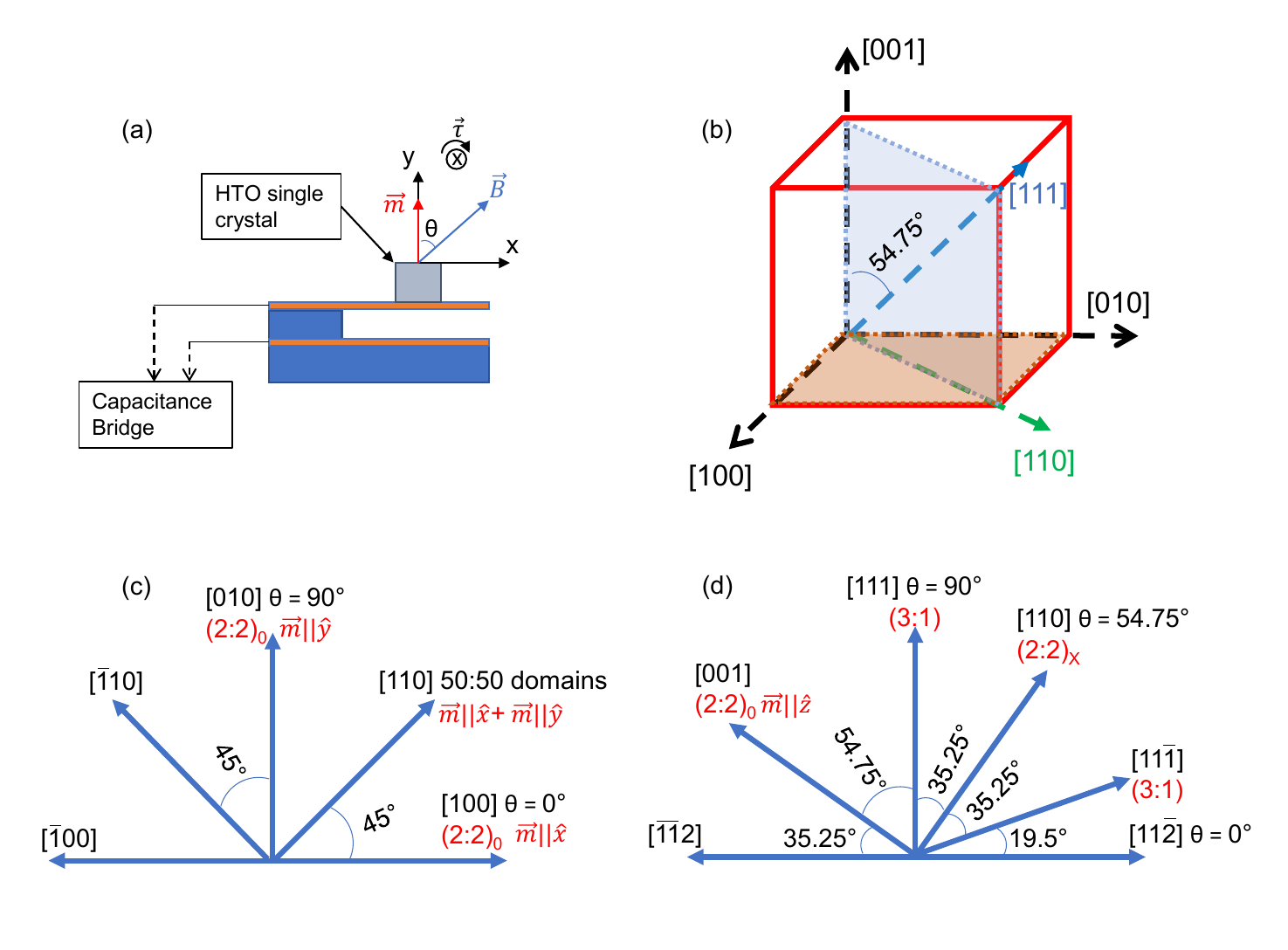}
\caption{\label{fig2}(a) Schematic diagram of the CTM set-up, which shows the sample placement on a flexible BeCu cantilever acting as the top plate of a parallel plate capacitor. A torque ($\vec{\tau}$) is generated when the magnetic moment ($\vec{m}$) is not aligned with the applied field ($\vec{B}$). The angle between $\vec{m}$ and $\vec{B}$ is indicated by $\theta$. (b) 3D representation of the cubic unit cell showing the two color-coded rotational planes of our interest, \hkl(001) (tan) and \hkl(1-10) (light blue). Rotational schematics for Ho$_2$Ti$_2$O$_7$ (HTO) single crystals. (c) \hkl(001) plane with magnetic field $\vec{B}$ rotating through the \hkl[100]$\rightarrow$\hkl[110]$\rightarrow$\hkl[010]$\rightarrow$\hkl[-110]$\rightarrow$\hkl[-100] directions. (d) \hkl(1-10) plane with $\vec{B}$ rotating through the \hkl[11-2]$\rightarrow$\hkl[11-1]$\rightarrow$\hkl[110]$\rightarrow$\hkl[111]$\rightarrow$\hkl[001] directions. Angular locations for important crystallographic directions and associated magnetic phases are included for both planes.}
\end{figure*}

\begin{figure*}[h!]
\centering
\includegraphics[width=0.9\textwidth]{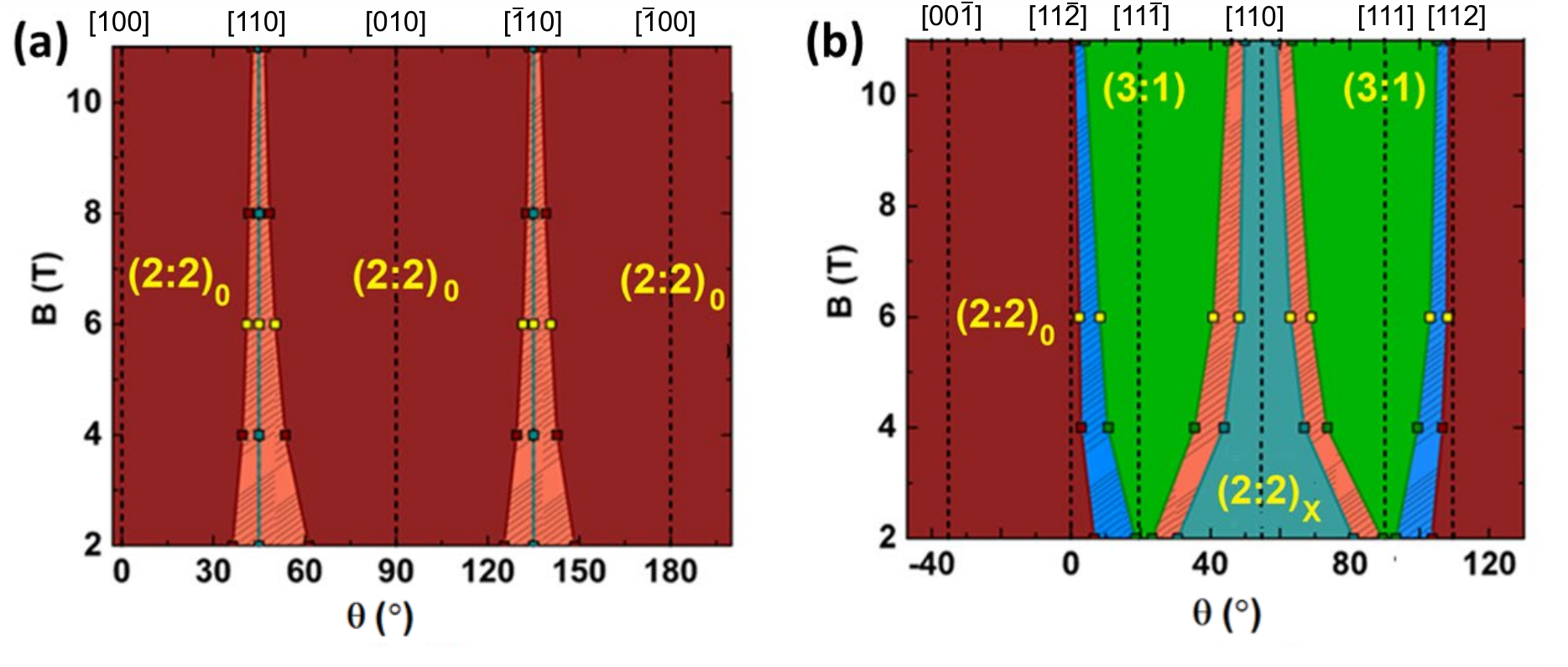}
\caption{\label{figPD}Angular-Field phase diagram, for applied fields ranging from 2-11~T while $\vec{B}$ rotates within the (a) \hkl(001) plane and (b) the \hkl(1-10) plane. The (2:2)$_0$, (2:2)$_X$, and (3:1) phases are indicated by the wine, cyan, and green colors, respectively. The blue and orange hatched regions indicate the angular ranges of domain states that form when the system transitions between stable phases. Crystallographic directions are indicated with vertical dashed lines.  }
\end{figure*}

\noindent Supplementary Figure~\ref{fig2} shows the schematic of the capacitive torque magnetometry (CTM) set-up and a 3D representation of the rotation planes. It also highlights the magnetic phases that we experimentally observe in both rotational planes when the magnetic field is aligned with the various high symmetry axes that lie on these planes. Based on the rotation schematics shown in Supplementary Figure~\ref{fig2}(c), the measured magnetic torque component is given as
\begin{equation}\label{eqnS1}
\tau_{n} = \hat{n}\cdot\vec{\tau} = [0, 0, \bar{1}]\cdot(\vec{m}\times\vec{B})\\ \textrm{with} ~ ~ ~ \vec{B} = B[\cos \theta, \sin \theta, 0]
\end{equation}
with $B$ the strength of the applied field. When rotating in the \hkl(001) plane, the zero crossings are observed at intervals of 45\degree \ and sharp turnovers in the data occur near any of the \hkl<110> family of directions. The system always remains in one of the Pauling states in this plane, in particular (2:2)$_0$ (or $Q~=~0$ \cite{Harrisbramwell}), with the field direction determining which of the 6 equivalent Pauling states dominates. These states are characterized by a saturated moment pointing along either the $\hat{x}$ or $\hat{y}$ or $\hat{z}$ directions. For our chosen plane of rotation, it is either along $\hat{x}$ or $\hat{y}$. We observe a sharp turnover in the data with a zero torque crossover points at or near the \hkl<110> directions. This is when all $\beta$-spins flip and $\vec{m}$ sharply rotates by 90$^{\circ}$. The intermediate phases that occur near the \hkl<110> directions could be understood as an equal volume fraction mixture of (2:2)$_0$ phases with half of the sample having $\vec{m}$ $\parallel \hat{x}$ and other half of the sample having $\vec{m}$ $\parallel \hat{y}$. This would result into a measured saturation magnetization of about 4.1 $\mu_B$/ Ho$^{3+}$ which is same as for the (2:2)$_X$ phase. However, in this rotation plane, we do not see any evidence of a stable (2:2)$_X$ phase as that would require a long range ordering among the $\beta$-spins. %As explained in the main text, the energy landscape of the system in this plane makes the (2:2)$_X$ phase inaccessible. 
% This transitions only involves spin flips on the $\beta$ spin sub-lattice (see main text). %Assuming that the four $\beta$ spins flip sequentially this yields intermediate spin textures, which have a well-defined torque response as a function of the angle of the field.   

%\noindent If the system starts in the $Q_x = 0$ then after one spin flip on the $\beta$ spin-sublattice and assuming that all spins on this sublattice are equally likely to flip, the average vector representing the magnetic moment in the system can be expressed as %(Fig.~\ref{fig1}B) 
%\;\;\;\;\;\;\;\;\;\;\;\;\;\;\;

%\noindent  $\vec{m}$ = $\sum_{i=1}^{16}\vec{m_{i}}=\frac{\mu }{\sqrt{3}} (14, 2, 0)$; \;\;\;\;\;\;\;\;\;  $\vec{\tau}$ = $\frac{16\mu B_{0}}{\sqrt{3}}(\frac{\sin \theta}{8}, - \frac{\cos \theta}{8}, \frac{7 \sin \theta}{8}- \frac{\cos \theta}{8})$ \\
%\noindent \bm {$\boxed {$\tau_{n}$ = $\hat{n}\cdot\vec{\tau}$ = $\frac{16\mu B_{0}}{\sqrt{3}}(\frac{-7 \sin \theta}{8}+ \frac{\cos \theta}{8})$}$} 

\begin{table*}[h!]
\begin{center}
\bgroup
\def\arraystretch{1.7}%
\caption{\label{tab:Table S1}Functional forms for calculated torque curves for the magnetic spin textures associated with the \hkl(001) rotational plane that are shown as solid/dotted curves in Figure 3 of the main text. Note, the magnetization, $\vec{m}$, is given as the magnetization per one unit cell, i.e., for 16 Ho$^{3+}$ sites, with $\mu$=10$\mu_B$ per site. Transitions between the indicated spin textures requires reversal of 4 $\beta$-spins, as indicated in red. }
\begin{tabular}{c c c c c} 
 \hline
 \hline
Phase & $\vec{B}$-direction &$\vec{m}$ = $\sum_{i=1}^{16}\vec{m_{i}}$& & $\tau_{n}$ = [0, 0, $\bar{1}$]$\cdot$($\vec{m}\times\vec{B}$) \\
\hline
(2:2)$_0$  & \hkl[100] &  $\frac{\mu }{\sqrt{3}}(16, 0, 0)$  \tikzmark{a}& &$\frac{-16\mu B}{\sqrt{3}}$ ${\sin} {\theta}$ \\\hline
50:50 domains  & \hkl<110> &$\frac{\mu }{\sqrt{3}}(8, 8, 0)$ \tikzmark{b} \tikzmark{c} & & $\frac{\mu B}{\sqrt{3}}(-8 \sin \theta+ 8 \cos \theta)$\\\hline
(2:2)$_0$  &\hkl[010] &  $\frac{\mu }{\sqrt{3}}(0, 16, 0)$ \tikzmark{d} & &$\frac{16\mu B}{\sqrt{3}}$ ${\cos} {\theta}$ \\\hline
\hline

\end{tabular}
\begin{tikzpicture}[overlay, remember picture, yshift=0.2\baselineskip, shorten >=.5pt, shorten <=.5pt,red]
    \draw [->] ({pic cs:a}) [bend left] to ({pic cs:b});
    \node[] (a) at (-4,-0.2) {\textcolor{red}{4x$\beta$}};
    \end{tikzpicture}
  \begin{tikzpicture}[overlay, remember picture, yshift=0.2\baselineskip, shorten >=.5pt, shorten <=.5pt, red]
    \draw [->] ({pic cs:c}) [bend left] to ({pic cs:d});
    \node[] (c) at (-4,-1.0) {\textcolor{red}{4x$\beta$}};
    \end{tikzpicture}  
\egroup
%\label{table:1}
\end{center}
\end{table*}

%In Fig.\ref{Fig2-1:data}(b) and (e), we compare the model calculated torque curves to the angular measurement in the $x$-$y$ plane and body diagonal plane taken at 6~T (black solid curves). The other solid-colored curves are associated with either the Pauling states or (3:1) spin configurations. The red or blue dashed curves in between are plotted to simulate the transitions between stable phases and are generated by assuming a changing volume fraction of spin flips on the beta or the alpha spin sub-lattices, respectively.\cite{SM} 

%As mentioned, in the \hkl(1-10) plane, there are phase transitions between the (2:2)$_0$ and (3:1) phases near the \hkl<112> directions, and between the (2:2)$_X$ and (3:1) states near the \hkl<110> directions.Starting with the first transition,

For each of the mentioned phases the torque response in a given applied field as a function of angle $\theta$, defined as the angle away from the \hkl[100] direction, can be calculated using Supplementary Equation \ref{eqnS1}. Supplementary Table \ref{tab:Table S1} shows the phases, the associated net magnetic moments, and the functional forms of the torque curves as $\vec{B}$ rotates in the \hkl(001) plane. Transitions between spin textures, which can involve spin flips on either the $\alpha$ or $\beta$ sublattice, lead to gradual changes in the value for $\vec{m}$~= $\frac{\mu}{\sqrt{3}}$ $\cdot~\vec{S}$, with $\mu$ = 10$\mu_B$ the moment per Ho$^{3+}$ ion and with $\vec{S}=(S_x,S_y,S_z)$. In the \hkl(001)  plane, transitions involve only $\beta$-spin flips and each flip leads to $\Delta S_x$ = $\pm$2 and $\Delta S_y$ = $\pm$2 while $S_z$ remains zero. The torque curves for these intermediate spin textures are calculated in the same way as the other curves, taking into account only this change in $\vec{S}$ (see Figure 3 of the main text for the calculated torque response as a function of {\bf $B(\theta)$}). The field-angle phase diagram for the \hkl(001) rotational plane as determined from the CTM measurements, is provided in Supplementary Figure \ref{figPD}(a).%Fig. 2 in the main text shows the comparison between the model calculated torque curves and the angular measurement taken at 6~T (black solid curves). The other solid-colored curves are associated with the Pauling states spin configurations. The red dashed curves in between are plotted to simulate the transitions between stable $Q~=~0$, (2:2)$_0$ phases and are generated by assuming a changing volume fraction of spin flips on the beta spin sub-lattices.   

\section{The \hkl(1-10) rotational plane}

In the plane that contains the body diagonal \hkl[111],
a similar procedure is used to calculate the angular dependence of the torque response. For this plane, shown in Supplementary Figure \ref{fig2} (d) and with $B$~$\geq$~2~ T, the torque response shows additional zero crossings and turnovers with significant changes in the observed torque amplitude when approaching different high symmetry axes.
%The calculation assumes that the metamagnetic phase transition has occurred along $<$111$>$ family of directions and 3-in/1-out phase has fully developed. 
Note, in this rotational plane $\alpha$-spins are involved in the phase transition between (2:2)$_0$  and the (3:1) monopole phases and $\beta$-spins are involved in the phase transition between the (2:2)$_X$  and (3:1) monopole phases.  We define 
\begin{subequations}
\begin{equation} 
\frac{[1, \bar{1}, 0]}{\sqrt{2}} = \hat{i}^{'}; \;\;\; \frac{[1, 1, \bar{2}]}{\sqrt{6}} = \hat{j}^{'}; \;\;\; \frac{[1, 1, 1]}{\sqrt{3}} = \hat{k}^{'}~(\textrm{with~} \hat{i}^{'}\times \hat{j}^{'}=\hat{k}^{'})
\end{equation}\begin{equation}
\vec{B} = B[\cos \theta~\hat{j}^{'} + \sin \theta~\hat{k}^{'}]\end{equation}
\begin{equation} \tau_{n} = \hat{n}\cdot\vec{\tau} = \frac{[\bar{1}, 1, 0]}{\sqrt{2}} \cdot(\vec{m}\times\vec{B})
\end{equation}\label{Bvectors}
\end{subequations}

\begin{table*}[h!]
\begin{center}
\bgroup
\def\arraystretch{1.7}%
\caption{\label{tab:Table S2} Functional forms for torque curves for the magnetic spin textures associated with the \hkl(1-10) rotational plane that are shown as solid curves in Figure 3 of the main text. Note, the magnetization, $\vec{m}$, is given as the magnetization per one unit cell, i.e., for 16 Ho$^{3+}$ sites, with $\mu$=10$\mu_B$ per site. Transitions between the indicated spin textures requires reversal of 4 $\alpha$- or $\beta$-spins, as indicated in blue and red text, respectively. }
\begin{tabular}{c c c c c} 
 \hline
 \hline
Phase & $\vec{B}$-direction &$\vec{m}$ = $\sum_{i=1}^{16}\vec{m_{i}}$ & & $\tau_{n}$ = $\frac{[\bar{1}, 1, 0]}{\sqrt{2}}$$\cdot$($\vec{m}\times\vec{B}$) \\
\hline
(2:2)$_0$ phase  & \hkl[00-1] &  $\frac{\mu }{\sqrt{3}}(0, 0, -16)$ \tikzmark{aa}& & $\frac{\mu B}{6} (-32 \cos \theta - 32\sqrt{2} \sin \theta)$ \\\hline
(3:1) monopole phase  & \hkl[11-1] &$\frac{\mu }{\sqrt{3}}(8, 8, -8)$ \tikzmark{ab} \tikzmark{ac}&  & $\frac{\mu B}{6}(16 \cos \theta - 32\sqrt{2} \sin \theta)$\\\hline
(2:2)$_X$ phase  & \hkl[110] &$\frac{\mu }{\sqrt{3}}(8, 8, 0)$ \tikzmark{ad} \tikzmark{e}&  & $\frac{\mu B}{6}(32 \cos \theta - 16\sqrt{2} \sin \theta)$\\\hline
(3:1) monopole phase   & \hkl[111] &$\frac{\mu }{\sqrt{3}}(8, 8, 8)$ \tikzmark{f} \tikzmark{g}  && $\frac{\mu B}{6}(48 \cos \theta)$\\\hline
(2:2)$_0$ phase  &\hkl[001] &  $\frac{\mu }{\sqrt{3}}(0, 0, 16)$ \tikzmark{h} &  & $\frac{\mu B}{6}(32 \cos \theta + 32\sqrt{2} \sin \theta)$ \\\hline
\hline

\end{tabular}
\begin{tikzpicture}[overlay, remember picture, yshift=0.2\baselineskip, shorten >=.5pt, shorten <=.5pt,blue]
\draw [->] ({pic cs:aa}) [bend left] to ({pic cs:ab});\node[] (aa) at (-4.7,0.5) {\textcolor{blue}{4x$\alpha$}};
\end{tikzpicture}
\begin{tikzpicture}[overlay, remember picture, yshift=0.2\baselineskip, shorten >=.5pt, shorten <=.5pt, red]
\draw [->] ({pic cs:ac}) [bend left] to ({pic cs:ad});\node[] (ac) at (-4.8,-0.3) {\textcolor{red}{4x$\beta$}};
\end{tikzpicture} 
\begin{tikzpicture}[overlay, remember picture, yshift=0.2\baselineskip, shorten >=.5pt, shorten <=.5pt, red]
\draw [->] ({pic cs:e}) [bend left] to ({pic cs:f});\node[] (e) at (-4.9,-1.0) {\textcolor{red}{4x$\beta$}};
\end{tikzpicture} 
\begin{tikzpicture}[overlay, remember picture, yshift=0.2\baselineskip, shorten >=.5pt, shorten <=.5pt,blue]
\draw [->] ({pic cs:g}) [bend left] to ({pic cs:h});\node[] (g) at (-5,-1.7) {\textcolor{blue}{4x$\alpha$}};
\end{tikzpicture}
\egroup
%\label{table:1}
\end{center}
\end{table*}

Supplementary Table~\ref{tab:Table S2} lists all the moment vectors and the functional forms of the angular dependence of the torque curves for the stable spin textures. When an $\alpha$-spin flips, each flip leads to $\Delta S_x$=$\Delta S_y$ = $\pm$2 and $\Delta S_z$ = $\pm$2. When a $\beta$-spin flips, each flip leads to $\Delta S_z$ = $\pm$2 while $S_x$ and $S_y$ remain unchanged. After every $\alpha$- or $\beta$-spin flip in a unit cell consisting of four tetrahedra, the system goes into a mixed domain phase where 25$\%$ of the system has made the phase transition (i.e., one tetrahedron in every unit cell has transitioned between a Pauling and a monopole state). Therefore, the Pauling:monopole volume fraction changes from 4:0 $\rightarrow$ 3:1 $\rightarrow$ 2:2 $\rightarrow$ 1:3 $\rightarrow$ 0:4 during a phase transition. This provides the appropriate scaling factors to extract the magnetic moment $\vec{m}$, and hence, the functional form of the torque equations for the intermediate phases during the phase transition. The torque curves for the ordered phases and these intermediate spin textures are shown in Figure 3 in the main text as solid and dotted curves, respectively. The field-angle phase diagram for the \hkl(1-10) rotational plane as determined from the CTM measurements, is provided in Supplementary Figure \ref{figPD}(b). %It also shows the comparison between the model calculated torque curves and the angular measurement taken at 6~T (black solid curves). The other solid-colored curves are associated with either the Pauling states or (3:1) spin configurations. The red or blue dashed curves in between are plotted to simulate the transitions between stable phases and are generated by assuming a changing volume fraction of spin flips on the beta or the alpha spin sub-lattices, respectively.
%In Fig.\ref{Fig2-1:data}(b) and (e), we compare the model calculated torque curves to the angular measurement in the $x$-$y$ plane and body diagonal plane taken at 6~T (black solid curves). The other solid-colored curves are associated with either the Pauling states or (3:1) spin configurations. The red or blue dashed curves in between are plotted to simulate the transitions between stable phases and are generated by assuming a changing volume fraction of spin flips on the beta or the alpha spin sub-lattices, respectively.\cite{SM} 

%As mentioned, in the \hkl(1-10) plane, there are phase transitions between the (2:2)$_0$ and (3:1) phases near the \hkl<112> directions, and between the (2:2)$_X$ and (3:1) states near the \hkl<110> directions.Starting with the first transition,

\begin{figure}
    \centering
    \includegraphics[width=0.9\textwidth]{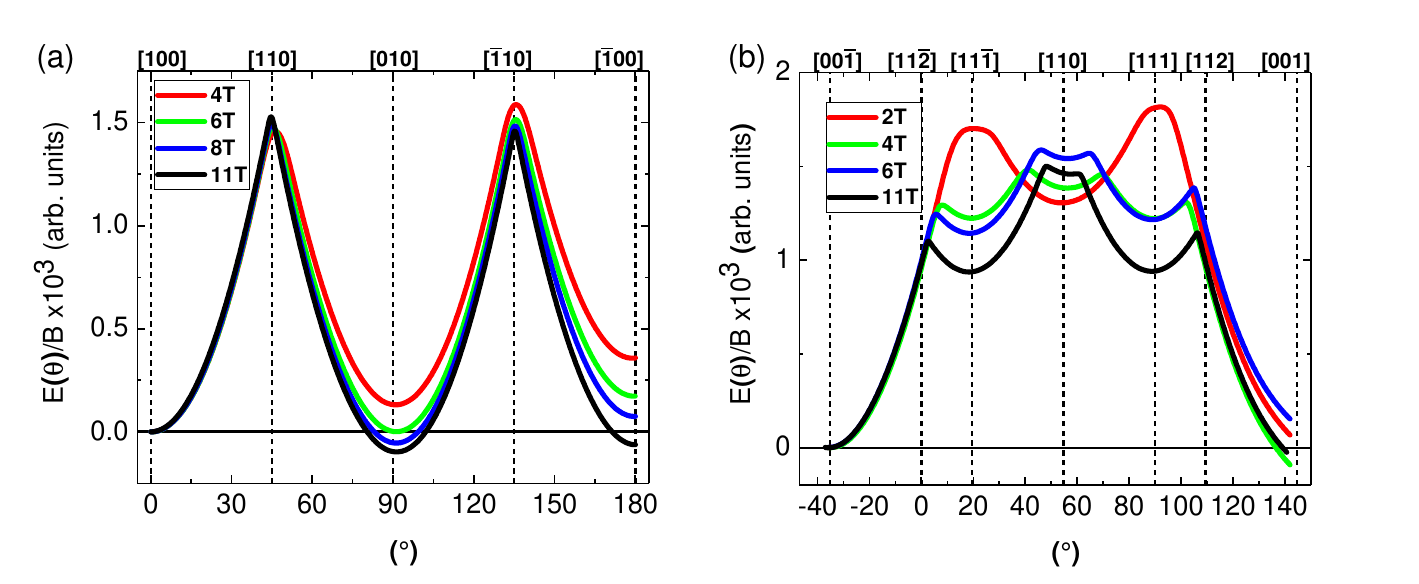}
    \caption{Field scaled integrated torque curves based on the curves shown in Figure 3 of the main manuscript. These curves represent contours on the magnetic anisotropy energy surface as a function of angle. The sample is rotated so that the field stays in (a) the \hkl(001) plane and (b) in the \hkl(1-10) plane. Crystallographic directions are indicated with vertical dashed lines. }
    \label{fig:MAE}
\end{figure}

\section{Integrated torque curve and the magnetic anisotropy energy}
From the measured torque response we can extract the anisotropy energy (E$_a$) as a function of field strength and direction. The magnetic torque is the derivative of this energy,
\begin{equation}
    \vec{\tau}=-\frac{dE_a}{d\theta}
\end{equation}
where $\theta$ is the rotation angle around a defined axis within the crystal. Measuring torque responses for different rotation planes thus provide insight into the shape of the anisotropy energy surface. In most systems this torque provides a direct means to determine the magneto-crystalline anisotropy, as most magnetic materials have a rather small crystal field that can be easily overcome by an applied field. Since, the spin-ice system has a very large crystal field splitting \cite{Opherden_2019}, rather than measuring the magneto-crystalline anisotropy surface, we are probing the energy surface associated with the various noncollinear spin textures that are stabilized under different magnetic field conditions. Integrating the torque curves reveal the relative energy differences between these possible spin textures as the sample is rotated in applied fields. Field-scaled integrated torque curves for both the \hkl(001) and \hkl(1-10) planes are plotted in Supplementary Figure \ref{fig:MAE} for various applied field strengths.   

We assume that the energy surface is continuous. Thus, it becomes clear that the (2:2)$_X$ phase, which is expected to form when the field is applied along any of the \hkl<110> directions, resides on a sharp energy maximum (i.e., is not stable) when rotating in the \hkl(001) plane, yet it resides on a local minimum in the energy surface when rotating in the \hkl(1-10) plane. It appears to reside on a saddle point in energy. Based on this energy surface, a long-range ordered (2:2)$_X$ phase does not appear to form when the field is confined to rotate in the \hkl(001) plane, yet an ordered (2:2)$_X$ phase does form when the field is confined to rotate in the \hkl(1-10) plane, with the angular range of this phase clearly showing a strong dependence on field-strength.

\section{The field sweep: phase transition between (3:1) and (2:2)$_X$}
For the field sweep measurement, a similar procedure is used to calculate the torque curves as a function of field strength. For these measurements, the field is misaligned away from the \hkl[111] direction by a small angle of $\Delta\theta$ $\approx$ 5\degree\ towards the \hkl[110] direction. 
We express the applied field as $\vec{B}$ = $B$[$\cos (\theta)$~$\hat{j}^{'}$ + $\sin (\theta)$~$\hat{k}^{'}$] and $\tau_{n} = \hat{n}\cdot\vec{\tau} = \frac{[\bar{1}, 1, 0]}{\sqrt{2}} \cdot(\vec{m}\times\vec{B})$, identical to the reference frame used for the \hkl(1-10) rotation. We note that $\theta$, defined as the angle between $\vec{B}$ and the \hkl[11-2] direction, is fixed during the measurement ($\theta$ = $90\degree$ - $\Delta\theta$), while $B$ varies between 7~T and -7~T. Supplementary Table~\ref{tab:Table S3} lists all the moment vectors and the torque curves for the stable spin textures. It also indicates which and how many spins per unit cell are flipped during a phase transition. When a $\beta$-spin flips, each flip leads to $\Delta S_z$ = $\pm$2, while $S_x$ and $S_y$ remain unchanged whereas, an $\alpha$-spin flip leads to $\Delta S_x$=$\Delta S_y$ = $\pm$2 with $S_z$ remaining unchanged. %This observation is based on the assumption that equally energetic spins flip at any given moment during a transition will occupy equal volume fractions in the sample. 

Assuming  $\Delta\theta$ $=$ 5\degree , the experimentally measured ratio of the slopes between the (3:1) phase and the (2:2)$_X$ phases gives a value of $\frac{\left|\vec{m}\right|_{sat}^{(3:1)}}{\left|\vec{m}\right|_{sat}^{(2:2)_X}}$ $\approx$ 1.18. This is extremely close to the expected value of $\frac{\left|\vec{m}\right|_{sat}^{(3:1)}}{\left|\vec{m}\right|_{sat}^{(2:2)_X}}$ = $\frac{5.0\mu_B}{4.1\mu_B}$ = 1.22 based on magnetization measurements reported by others\cite{KreyLegl,petrenko}. An agreement within 5 $\%$ between the expected and experimental values gives us a great deal of confidence in the measurement set-up and phenomenological model. We also recognize that there is no stable phase when the field reverses its orientation. This occurs between -0.1~T $\rightarrow$ -1.8~T and 0.1~T $\rightarrow$ 1.8~T when all $\alpha$-spins flip.  However, approaching the same region from the high field end ($\pm{7}$~T) results into a clean (2:2)$_X$ phase, which follows after the $\beta$-spins flip. Because different sets of spin-flips are involved during the low and high field transitions in the field sweep measurement, a hysteretic butterfly loop within the field range of -1.8 T$<B<$1.8~T results. The energy cost for a spin flip to occur and create the (3:1) state on a single tetrahedron is likely too high at low field, i.e., when the (3:1) state is inaccessible,  hysteresis appears, with the 8-$\alpha$ spins likely flipping in pairs.

\begin{table*}[h!]
\begin{center}
\bgroup
\def\arraystretch{1.9}
\caption{\label{tab:Table S3}Functional forms for torque curves for the magnetic spin textures associated with the field sweep measurements. For the solid curves in Figure 4(b) of the main text, with $\theta$ = 85\degree. The magnetization, $\vec{m}$, is given as the magnetization per one unit cell, i.e., for 16 Ho$^{3+}$ sites, with $\mu$=10$\mu_B$ per site. Transitions between the indicated spin textures requires reversal of 8 $\alpha$- or 4 $\beta$-spins, as indicated in blue and red text, respectively. }
\begin{tabular}{c c c c c} 
 \hline
 \hline
Field range (sweep direction) & Phase &$\vec{m}$ = $\sum_{i=1}^{16}\vec{m_{i}}$ & & $\tau_{n}$ = $\frac{[\bar{1}, 1, 0]}{\sqrt{2}}$$\cdot$($\vec{m}\times\vec{B}$) \\
\hline
3 T$<B<$ 7 T (Down) & (3:1) monopole  &  $\frac{\mu }{\sqrt{3}}(8, 8, 8)$ \tikzmark{aaa}& & $\frac{\mu B}{6} (48 \cos \theta )$ \\\hline
-0.1 T$<B<$ 1.8 T (Down) & (2:2)$_X$    &$\frac{\mu }{\sqrt{3}}(8, 8, 0)$ \tikzmark{aab} \tikzmark{aac}&  & $\frac{\mu B}{6}(32 \cos \theta - 16\sqrt{2} \sin \theta)$\\\hline
-1.8 T$<B<$ -0.1 T (Down) & No stable phase   &$\frac{\mu }{\sqrt{3}}(-8, -8, 0)$ \tikzmark{aad} \tikzmark{ae}&  & No stable phase \\\hline
-7 T$<B<$ -3 T (Down/Up) & (3:1) monopole   &$\frac{\mu }{\sqrt{3}}(-8, -8, -8)$ \tikzmark{af} \tikzmark{ag} && $\frac{\mu B}{6}(-48 \cos \theta)$\\\hline
-1.8 T$<B<$ 0.1 T (Up) &  (2:2)$_X$   &$\frac{\mu }{\sqrt{3}}(-8, -8, 0)$ \tikzmark{baad} \tikzmark{bae}&  & $\frac{\mu B}{6}(-32 \cos \theta + 16\sqrt{2} \sin \theta)$\\\hline
0.1 T$<B<$ 1.8 T (Up) & No stable phase  &$\frac{\mu }{\sqrt{3}}(8, 8, 0)$ \tikzmark{caad} \tikzmark{cae}&  & No stable phase \\\hline
3 T$<B<$ 7 T (Up) & (3:1) monopole   &  $\frac{\mu }{\sqrt{3}}(8, 8, 8)$ \tikzmark{baaa}& & $\frac{\mu B}{6} (48 \cos \theta )$ \\\hline
\hline
\end{tabular}
\begin{tikzpicture}[overlay, remember picture, yshift=0.2\baselineskip, shorten >=.5pt, shorten <=.5pt,red]
\draw [->] ({pic cs:aaa}) [bend left] to ({pic cs:aab});\node[] (aaa) at (-4.8,1.3) {\textcolor{red}{4x$\beta$}};
\end{tikzpicture}
\begin{tikzpicture}[overlay, remember picture, yshift=0.2\baselineskip, shorten >=.5pt, shorten <=.5pt, blue]
\draw [->] ({pic cs:aac}) [bend left] to ({pic cs:aad}); \node[] (aac) at (-4.8,0.5) {\textcolor{blue}{8x$\alpha$}};
\end{tikzpicture} 
\begin{tikzpicture}[overlay, remember picture, yshift=0.2\baselineskip, shorten >=.5pt, shorten <=.5pt, red]
\draw [->] ({pic cs:ae}) [bend left] to ({pic cs:af});\node[] (ae) at (-4.8,-0.3) {\textcolor{red}{4x$\beta$}};
\end{tikzpicture} 
\begin{tikzpicture}[overlay, remember picture, yshift=0.2\baselineskip, shorten >=.5pt, shorten <=.5pt, red]
\draw [->] ({pic cs:ag}) [bend left] to ({pic cs:baad});\node[] (aac) at (-5.0,-1.1) {\textcolor{red}{4x$\beta$}};
\end{tikzpicture} 
\begin{tikzpicture}[overlay, remember picture, yshift=0.2\baselineskip, shorten >=.5pt, shorten <=.5pt, blue]
\draw [->] ({pic cs:bae}) [bend left] to ({pic cs:caad});\node[] (aac) at (-5.2,-2.0) {\textcolor{blue}{8x$\alpha$}};
\end{tikzpicture} 
\begin{tikzpicture}[overlay, remember picture, yshift=0.2\baselineskip, shorten >=.5pt, shorten <=.5pt, red]
\draw [->] ({pic cs:cae}) [bend left] to ({pic cs:baaa});\node[] (aac) at (-5.3,-2.9) {\textcolor{red}{4x$\beta$}};
\end{tikzpicture} 
 
     %\begin{tikzpicture}[overlay, remember picture, yshift=0.2\baselineskip, shorten >=.5pt, shorten <=.5pt,blue]
    %\draw [->] ({pic cs:ag}) [bend left] to ({pic cs:ah});
    %\node[] (ag) at (-5,-1.7) {\textcolor{blue}{4x$\alpha$}};
    %\end{tikzpicture}
\egroup
%\label{table:1}
\end{center}
\end{table*}

\begin{figure*}[h!]
\centering
\includegraphics[width=\textwidth]{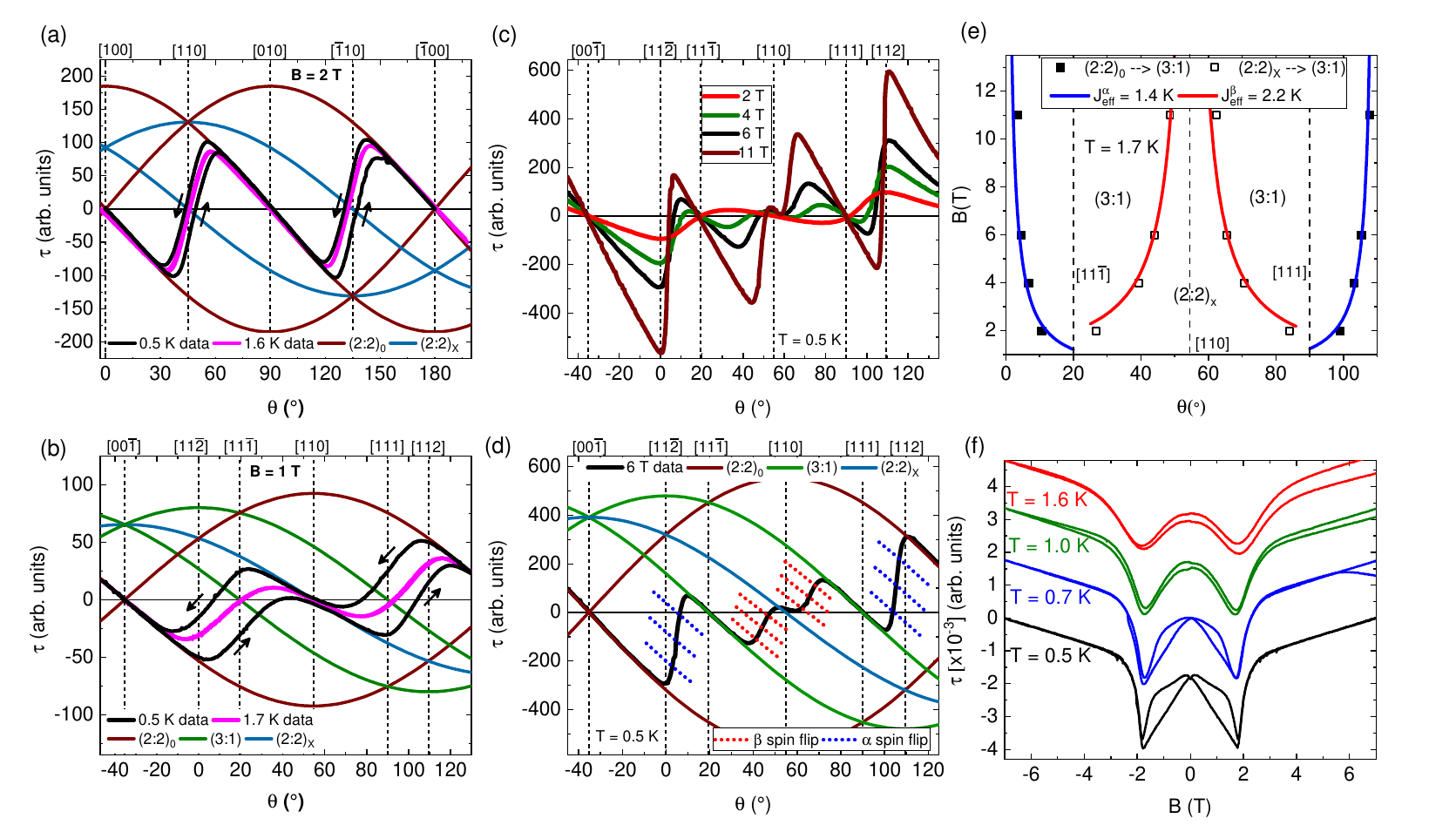}
\caption{\label{fig-hysteresis}(a) The \hkl(001) rotational plane: CTM measurement showing torque as a function of angle at $B$ = 2~T, measured at \textit{T} = 0.5~K (solid black curve) and \textit{T} = 1.6~K (solid magenta curve). The arrows indicate the sweep directions. The wine and blue solid lines are calculated torque curves associated with the (2:2)$_0$ and (2:2)$_X$ phases, respectively (see Supplementary Table \ref{tab:Table S1}). (b) The \hkl(1-10) rotational plane: CTM measurement showing torque as a function of angle at $B$ = 1~T, measured at \textit{T} = 0.5~K (solid black curve) and \textit{T} = 1.7~K (solid magenta curve). The arrows indicate the sweep directions. The wine, blue and green solid lines are calculated torque curves associated with the (2:2)$_0$, (2:2)$_X$, and (3:1) monopole phases, respectively (see Supplementary Table \ref{tab:Table S2}). (c) CTM response at $T$ = 1.7~K when $\vec{B}$ rotates in the plane containing the body diagonal (the \hkl[11-2] direction corresponds to 0\degree\ ). (d) CTM response in a 6~T applied field for the \hkl(1-10) plane at $T$ = 1.7~K. The wine, blue and green solid lines are calculated torque curves associated with the (2:2)$_0$, (2:2)$_X$, and (3:1) monopole phases, respectively (see Supplementary Table \ref{tab:Table S2}). (e) Critical angles as a function of applied field for the \hkl(1-10) plane rotation data at 1.7~K for transitions between (2:2)$_0$ $\Leftrightarrow$ (3:1) (solid squares) as well as (2:2)$_X$ $\Leftrightarrow$ (3:1) (open squares). The blue and red curves correspond to model curves using $J^{\alpha}_{eff}$ = 1.4~K and $J^{\beta}_{eff}$ = 2.2~K, respectively. Crystallographic directions are indicated by vertical dashed lines. (f) Capacitance change as a function of applied field, misaligned away from the \hkl[111] direction at $T$ = 0.5~K (black), $T$ = 0.7~K (blue), $T$ = 1.0~K (green), and $T$ = 1.6~K (red). The curves are vertically translated for clarity. }
\end{figure*}

\section{Temperature dependent hysteresis in torque magnetometry}

\subsection{The \hkl(001) rotational plane}
As shown in Figure 3 (a) in the main text, the steep transition around the \hkl<110> family of directions develops obvious hysteresis at low applied fields at $T$ = 500~mK. The hysteresis is a clear sign of the glassiness of the system. The field rotation speed may be important in determining the degree of hysteresis, as the system will be effectively out of equilibrium if the field rotates too fast. Here we note that hysteresis is also present at higher fields, but gradually closes with increasing field. We have also investigated how the hysteresis changes with increasing temperature (for $B$ = 2~T), Supplementary Figure \ref{fig-hysteresis}(a) shows the torque response at both $T$ = 500~mK as well as $T$ = 1.6~K. The wine and blue solid lines correspond to calculated torque curves associated with the (2:2)$_0$ and (2:2)$_X$ phases, respectively (see Supplementary Table \ref{tab:Table S1}). It is clear that the hysteresis disappears with increasing temperature. Therefore, we associate the appearance of the hysteresis with the local internal fields (spin-spin interactions) of the spin-ice state as it appears below the spin freezing temperature.  %We extract the energy scale associated with this internal field, and thus the spin-spin interaction strength, in the main text.

%\begin{figure*}
 %   \begin{tikzpicture}
  %  \node(a){\includegraphics[width=\textwidth]{Field_sweep-2.png}};
   % \node at (a.north)
    %[
    %anchor=center,
 %   xshift=23mm,
 %   yshift=-22mm
 %   ]
 %   {
  %      \includegraphics[width=0.25\textwidth]{Figures/3I-1O.png}
 %   };
%    \node at (a.east)
 %   [
 %   anchor=center,
  %  xshift=-23mm,
 %   yshift=17mm
 %   ]
%    {
 %       \includegraphics[width=0.25\textwidth]{Figures/beta1.png}
 %   };
 %    \node at (a.east)
 %   [
  %  anchor=center,
  %  xshift=-23mm,
 %   yshift=-19mm
 %   ]
 %   {
 %       \includegraphics[width=0.25\textwidth]{Figures/beta2.png}
  %  };
  %   \node at (a.west)
  %  [
  %  anchor=center,
  %  xshift=43mm,
  %  yshift=15mm
  %  ]
  %  {
   %     \includegraphics[width=0.25\textwidth]{Figures/alpha1.png}
 %   };
 %    \node at (a.west)
%    [
%    anchor=center,
%    xshift=43mm,
%    yshift=-19mm
 %   ]
 %   {
 %       \includegraphics[width=0.25\textwidth]{Figures/alpha2.png}
 %   };
%    \end{tikzpicture}
%    \caption{(Color online) Capacitance change as a function of applied field at T = 0.5 K, the black arrows indicate the sweep direction. The solid lines are volume-scaled calculated torque curves for the 3-out/1-in phase (green) and the Q = X phase (dark cyan). The dotted lines are the volume-scaled calculated torque curves after spin flips have occurred in the system (red, $\beta$ spin flips and blue, $\alpha$ spin flips). All inset figures show 4 spins within a tetrahedron in xyz-coordinate system. It identifies the specific spin sublattice being flipped as external field sweeps between +/- 7 tesla. 
 %   }
%\end{figure*}

\subsection{The \hkl(1-10) rotational plane}
As mentioned in the main text and as shown in Supplementary Figure \ref{fig-hysteresis}(b), the transitions around the \hkl[112] and \hkl[11-2] directions between the (2:2)$_0$ and  (3:1) monopole phase disappear below $B$~=~2~T, i.e., the (3:1) monopole phase does not form at low field. As a result hysteresis appears around the \hkl[111]  and \hkl[11-1] directions for rotations in $B$~=~1~T. This hysteresis is clearly associated with the spin-ice phase as it has fully disappeared at T = 1.7~K (the solid magenta curve in Supplementary Figure \ref{fig-hysteresis}(b)). The wine, blue, and green solid lines correspond to calculated torque curves associated with the (2:2)$_0$, (2:2)$_X$ and (3:1) monopole phases, respectively (see Supplementary Table \ref{tab:Table S2}). We associate the appearance of the hysteresis with the local internal fields (spin-spin interactions) of the spin-ice state and assume that in this rotational plane it also appears below the spin freezing temperature. 

In Supplementary Figure \ref{fig-hysteresis}(c) we show the CTM response measured at 1.7~K in various applied fields. Using the same procedure as described in the main text we compare these angular sweeps with the calculated torque curves and extract the critical angles for the (2:2)$_0$ $\Leftrightarrow$ (3:1) and (2:2)$_X$ $\Leftrightarrow$ (3:1) transitions (see Supplementary Figure \ref{fig-hysteresis}(d)). The critical angles as a function of applied field are plotted in Supplementary Figure \ref{fig-hysteresis}(e). The thermal smearing of the transitions and the fact that at 2~T we do not observe the (3:1) state, makes the analysis as described in the main text difficult for the high temperature data. Instead of fitting the critical angles, we plot them together with the functional forms for $B(\theta)$ using 1.4 and 2.2~K for $J^{\alpha}_{eff}$ and $J^{\beta}_{eff}$, respectively. The agreement between $B(\theta)$ and the data is very good. %We extract the energy scale associated with the spin-spin interaction strength $J_{eff}$, as in the main text, and the results are shown in Fig. \ref{fig-hysteresis}(d). These extracted value differ slightly from T = 0.5 K results reported in the main text. However, we clearly see two different MCPs associated with two different transitions.

%\begin{figure*}[h!]
%\centering
%\includegraphics[width=\textwidth]{Compressed_Crystalline_Anisotropy_Energy_Updated_4_15_20.png}
%\caption{\label{fig-energy}(Color online) Anisotropy energy surface}
%\end{figure*}

\subsection{Temperature dependent hysteresis in field sweep with $\vec{B}$ near the \hkl[111] direction}
 We have also investigated how the field sweep changes as the temperature is gradually increased from $T$ = 500~mK to $T$ = 1.6~K (see Supplementary Figure \ref{fig-hysteresis}(f)). For these measurements, the field is misaligned away from the \hkl[111] direction by a small angle of $\Delta\theta$ $\approx$ 5\degree\ towards the \hkl[110] direction. While the linear responses at high fields ($B >$ 2~T) and at low fields (0~T $< B <$ 1.5~T) remain, the sharp turnover around $B$ = 1.8~T gradually smears out at higher temperature. The hysteresis around zero field disappears fully when the sample is heated to above $T$ = 0.7~K, i.e., around the spin freezing temperature. We observe no hysteresis around the sharp feature at $B$ = 1.8~T, which is expected and consistent with reports by others \cite{KreyLegl}. We also notice that as the temperature is increased, the value for $\Delta C$ at the turnover ($B$ = 1.8~T) decreases in amplitude. This is consistent with the formation of thermal defects in the (2:2)$_X$ phase (with $\vec{S}~=(8,8,0)$) resulting in a gradual increase in $S_z$ as the long-range antiferromagnetic order on the $\beta$-spins gets randomized. Additionally, there is a drift in the measurement at 1.6~K (a little drift is observed at 1~K as well) and we associate this trend with a systematic drift in the sample temperature during field sweep. So, there is a clear evidence that the capacitance picks up an additional term because of the temperature variation during the field sweep sequence. This is expected in a sorption-pump based He3 cooling system because higher field burns helium at a faster rate. 
 
 \section{Hamiltonian and Monte Carlo Simulations}

The Hamiltonian of interest for Ho$_2$Ti$_2$O$_7$ is,
\begin{equation}
        H = -J_{1} \sum_{\langle i,j \rangle} \tilde{\textbf{S}}_{i} \cdot \tilde{\textbf{S}}_{j}  -J_{2} \sum_{\langle \langle i,j \rangle \rangle} \tilde{\textbf{S}}_{i} \cdot \tilde{\textbf{S}}_{j}  
  + D r^{3}_{nn} \sum_{i>j} \Big(\frac{\tilde{\textbf{S}}_{i} \cdot \tilde{\textbf{S}}_{j}}{|\textbf{r}_{ij}|^3} - \frac{ 3 (\tilde{\textbf{S}}_{i} \cdot \textbf{r}_{ij}) (\tilde{\textbf{S}}_{j} \cdot \textbf{r}_{ij})}{|\textbf{r}_{ij}|^5} \Big)
 - g \mu_B \sum_{i} \vec{B} \cdot \tilde{\textbf{S}_i} 
\label{eq:Ham}
\end{equation}
where $\tilde{\textbf{S}}_{i}$ are classical spin vectors with $|\tilde{\textbf{S}}_i|=1$. The tilde is used to indicate that the spins are constrained to point along the local \hkl<111> axis of the tetrahedra they belong to. $\textbf{r}_{i}$ is the real-space location of site $i$, $\textbf{r}_{ij} \equiv \textbf{r}_{i} - \textbf{r}_{j}$, $\langle i,j \rangle$ ($\langle \langle i,j \rangle \rangle$) refers to nearest-neighbor (next nearest-neighbor) bonds, $r_{nn}$ is the nearest-neighbor bond distance, $J_1$ ($J_2$) is the nearest-neighbor (next-nearest neighbor) interaction strength, and $D$ is the strength of the long-range dipolar term. The $i>j$ notation guarantees each of pair of spins is only counted once.
$g \mu_B$ is the size of the magnetic moment and  $\vec{B}$ is the applied magnetic field.

The spins are effectively Ising-like since they are constrained to be aligned along a local \hkl<111> axis, i.e.,
\begin{equation}
        \tilde{\textbf{S}}_i = \sigma_{i} z^{a(i)}
\label{eq:Sz}
\end{equation}
where $\sigma_{i} = \pm 1$ is a scalar and  $z^{a(i)}$ is a unit vector depending on the sublattice $a=0,1,2,3$ to which the site $i$ belongs. The directions of the $z^a$ vectors are shown in Supplementary Table \ref{T4}.
\begin{table*}[htpb]
\begin{center}
\caption{Sublattice unit vectors}\label{T4}
\begin{tabular}{|c|c|}
\hline
$a$ (sublattice)   &  $z^a = (x,y,z)$   \tabularnewline
\hline
0               &  $\frac{1}{\sqrt{3}} (+1,+1,+1)$\tabularnewline
1               &  $\frac{1}{\sqrt{3}} (+1,-1,-1)$\tabularnewline
2               &  $\frac{1}{\sqrt{3}} (-1,+1,-1)$\tabularnewline
3               &  $\frac{1}{\sqrt{3}} (-1,-1,+1)$\tabularnewline
\hline
\end{tabular}
\end{center}
\end{table*}

When the dipolar interaction is truncated to the nearest neighbor term, the Hamiltonian becomes,
\begin{equation}
        H_{NN} =  J^{s-DSI}_{1,eff} \sum_{\langle i,j \rangle} \sigma_i \sigma_j - g \mu_B \sum_{i} \sigma_i \vec{B} \cdot z^{a(i)}
\label{eq:NNHamlocalz}
\end{equation}
where $J^{s-DSI}_{1,eff} \equiv  \Big(\frac{J_{1}+ 5 D}{3}\Big)$. For the case of zero applied field, $J^{s-DSI}_{1,eff} > 0 $ results in the "2 in-2 out ice rule" for the ground state manifold of the spin ice.

We perform classical Monte Carlo (MC) simulations for both the nearest-neighbor (NN) and the standard dipolar spin-ice (s-DSI) models, with the parameters shown in Supplementary Table \ref{T5}. The s-DSI model includes long-range dipolar interactions via the Ewald summation method~\cite{DeLeeuw_dipolar, RogerGMelko, Wang_Holm, Jaubert_thesis}, the parameter set for the s-DSI model is taken from Refs.\cite{Bramwell_2020,Bramwell2001SpinMaterials}

\begin{table*}[htpb]
\begin{center}
\caption{The Parameter values for $J_1$, $D$, and $g$ used in the MC simulations for the NN and s-DSI model, respectively. The values for $J^{s-DSI}_{1,eff}$ are provided as a reference, but they are not used in the simulations. }\label{T5}
\begin{tabular}{|c|c|c|c|c|}
\hline
Model                   &  $J_1$  (K) & $D$ (K)   & $J_{1,eff}= \frac{J_1+5D}{3}$ (K)       &  $g$ (dimensionless) \tabularnewline
\hline
NN   &  +5.40     & 0       &   1.80 				     & 10 \tabularnewline
s-DSI (NN dipolar)      &  -1.56     & 1.41    &   1.83     			     & 10 \tabularnewline
\hline
\end{tabular}
\end{center}
\end{table*}

In Supplementary Table \ref{T6} we compare our values to others reported for HTO and Dy$_2$Ti$_2$O$_7$ (DTO). 
\begin{table}[htpb]
\begin{center}
\caption{The table summarizes values for $J_1$, $J_2$, $J_3$, $D$, and $J^{s-DSI}_{1,eff}$ as reported by others\cite{Henelius,Samarakoon,Ruff,Bramhartog}. $J_{3a}$ and $J_{3b}$ refer to third nearest neighbors of a and b types, respectively (see refs. \cite{Henelius,borzi} ).}\label{T6}
\begin{tabular}{|c|c|c|c|c|c|}
\hline
Ref                   &  $J_1$ (K) &$J_2$ (K) &  $J_{3a}$ ($J_{3b}$) (K)  & $D$ (K)    & $J_{1,eff}$  = $\frac{J_1+5D}{3}$ (K)      \tabularnewline
\hline
DTO\cite{Henelius}   &   3.41             &-0.14         &   0.025 (0.025) & 1.32    	&1.06			    \tabularnewline
  DTO\cite{Samarakoon}   &   3.41             &0         &   0.014 (0.096) &  1.32    	& 1.06			    \tabularnewline
 DTO\cite{Ruff}   &   -3.72             &  -        &   -0.02 &  1.41  & 1.11 				    \tabularnewline
HTO\cite{Ruff}   &   -1.65             & -        &  -  &  1.41    		&  1.80		    \tabularnewline
HTO\cite{Bramhartog}   &   -1.56             & -        &  -  &  1.41    		&  1.83		    \tabularnewline
\hline
\end{tabular}
\end{center}
\end{table}
\begin{figure}
    \centering
    \includegraphics[width=1.0\textwidth]{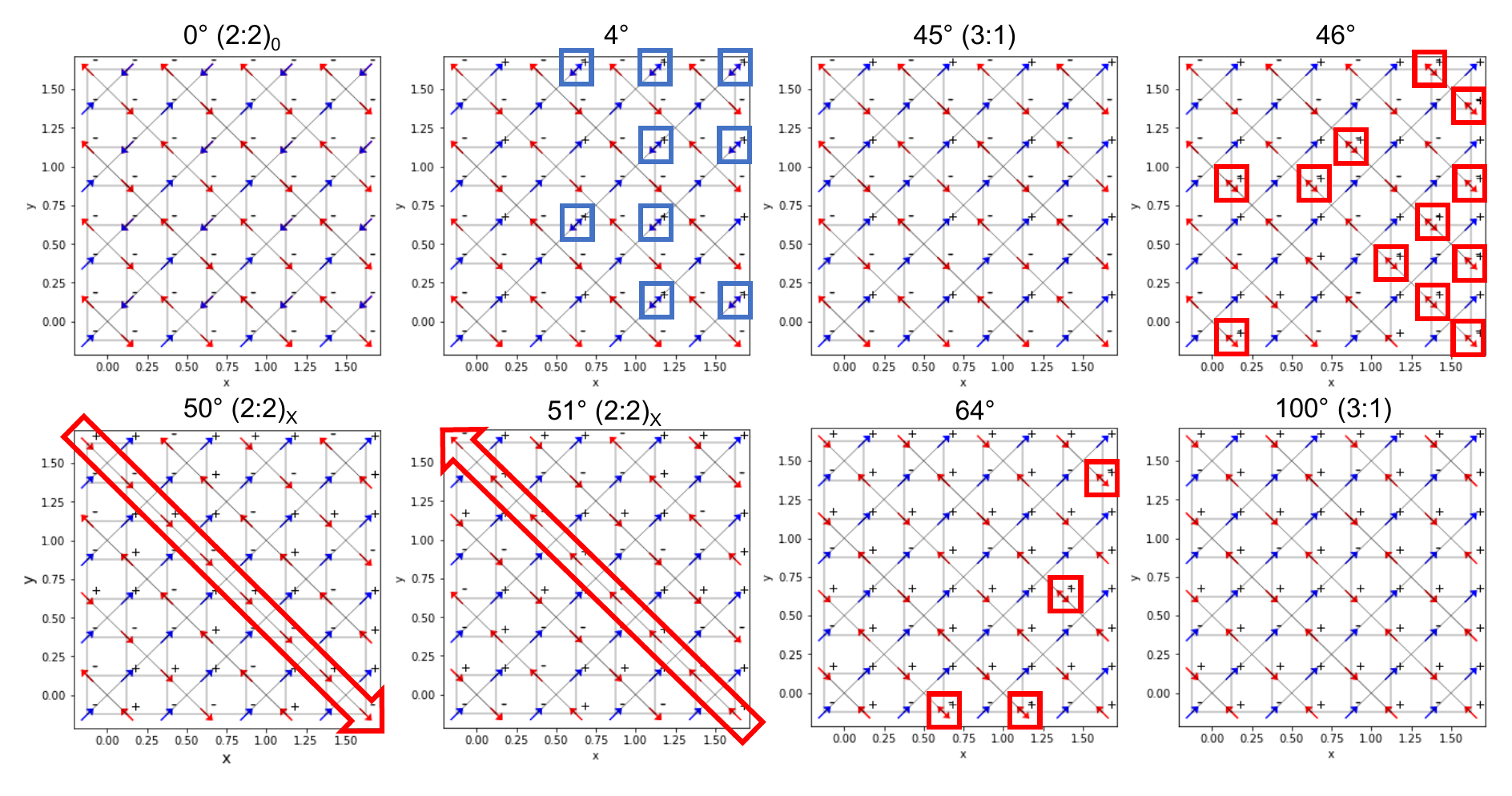}
    \caption{Snapshots of the spin configurations recorded for various angles between the applied field and the \hkl[001] direction. The snapshots are 2-dimensional projections of the 2x2x2 unit cells, projected down the $z$-axis ($\beta$-spins are red and the $\alpha$-spins are blue). The stable phases are indicated, (2:2)$_0$ at 0\degree (when $\vec{B}$ is along the [001]), (3:1) at 45\degree and 100\degree, and (2:2)$_X$ at 50\degree and 51\degree. The open red arrows highlight the flipped directions of the $\beta$-spin chains. At the intermediate angles of 4\degree, 46\degree, and 64\degree, the spin textures adopt an intermediate state in which the two stable phases involved in the transition form a domain structure. The blue and red boxes highlight the spin columns that host unflipped $\alpha$- and $\beta$-spins, respectively. }
    \label{fig:MCconf}
\end{figure}

We simulate finite size pyrochlore clusters with periodic boundary conditions with $N_{spins} = 16 \times 4^3 = 1024$ lattice sites and employ
the Ewald summation method~\cite{DeLeeuw_dipolar, RogerGMelko, Wang_Holm, Jaubert_thesis}. Demagnetization effects~\cite{RogerGMelko}
are found to be significant especially for field directions close to the observed phase boundaries. We model this with the help of a surface term in the Ewald summation~\cite{Wang_Holm}, $U^{surf} = D r_{nn}^3 \frac{2\pi}{(2\mu^{'}+1) N_{spins}} \sum_{i,j} \tilde{\bf{S}}_i \cdot \tilde{\bf{S}}_j $.
Incorporating this term with $\mu'=1$ provides a good match to the experimental observations.

Our MC sampling procedure employs the standard accept-reject Metropolis algorithm with an equal probability of ``loop'' moves
(effective for sampling the low energy spin-ice configurations at low temperature)
~\cite{Melko_Hertog_Gingras, RogerGMelko, Barkema_Newman}, and single spin flip moves (effective at high temperature). We measure the thermal average of the magnetization $\langle \vec{m} \rangle$ (per unit cell) using $10^5$ MC samples. One sample per MC sweep is recorded for thermal averaging, and one MC sweep comprises of a set of $N_{spins}$ proposed moves (which are individually of loop or single spin flip type).
We then compute the component of the torque along $\hat{n}$ (per unit cell) $\tau_n \equiv \hat{n} \cdot (\langle \vec{m} \rangle \times \vec{B})$, as discussed in the previous section. This quantity corresponds directly to what is being measured in the torque magnetometry experiment (up to device specific constants). We scan the space of field alignment angles in steps of 1\degree (or smaller near phase boundaries), considering the same range of angles as that recorded in the experiment.

In Supplementary Figure \ref{fig:MCconf} we show snapshots of spin textures recorded during the MC simulations at $T$ = 0.5~K as the field was rotated from the [001] direction at 0\degree. We show snapshots at various relevant angles, observing the expected stable phases (2:2)$_0$ at 0\degree, (3:1) at 45\degree and 100\degree, and (2:2)$_X$ at 50\degree and 51\degree. The reason that two (2:2)$_X$ configurations are shown is because in the  47 - 63\degree range, the spin texture toggles between the two possible states with oppositely oriented $\beta$-spin chains every couple of degrees, indicating that these textures are indeed degenerate. No defects are observed in either of the (2:2)$_X$ configurations, suggesting that thermal fluctuations at $T$ = 0.5~K are small and the state is relatively stable. Lastly, the configurations taken at angles of 4\degree, 46\degree, and 64\degree show intermediate or transient spin textures in which the two stable phases involved in the transition form a domain structure. This is visible from the presence of the double arrows, which indicates that in that column of spins, both in and out spins (either unflipped $\alpha$- or $\beta$- spins) exist.%. .  

\begin{figure}
    \centering
    \includegraphics[width=0.5\textwidth]{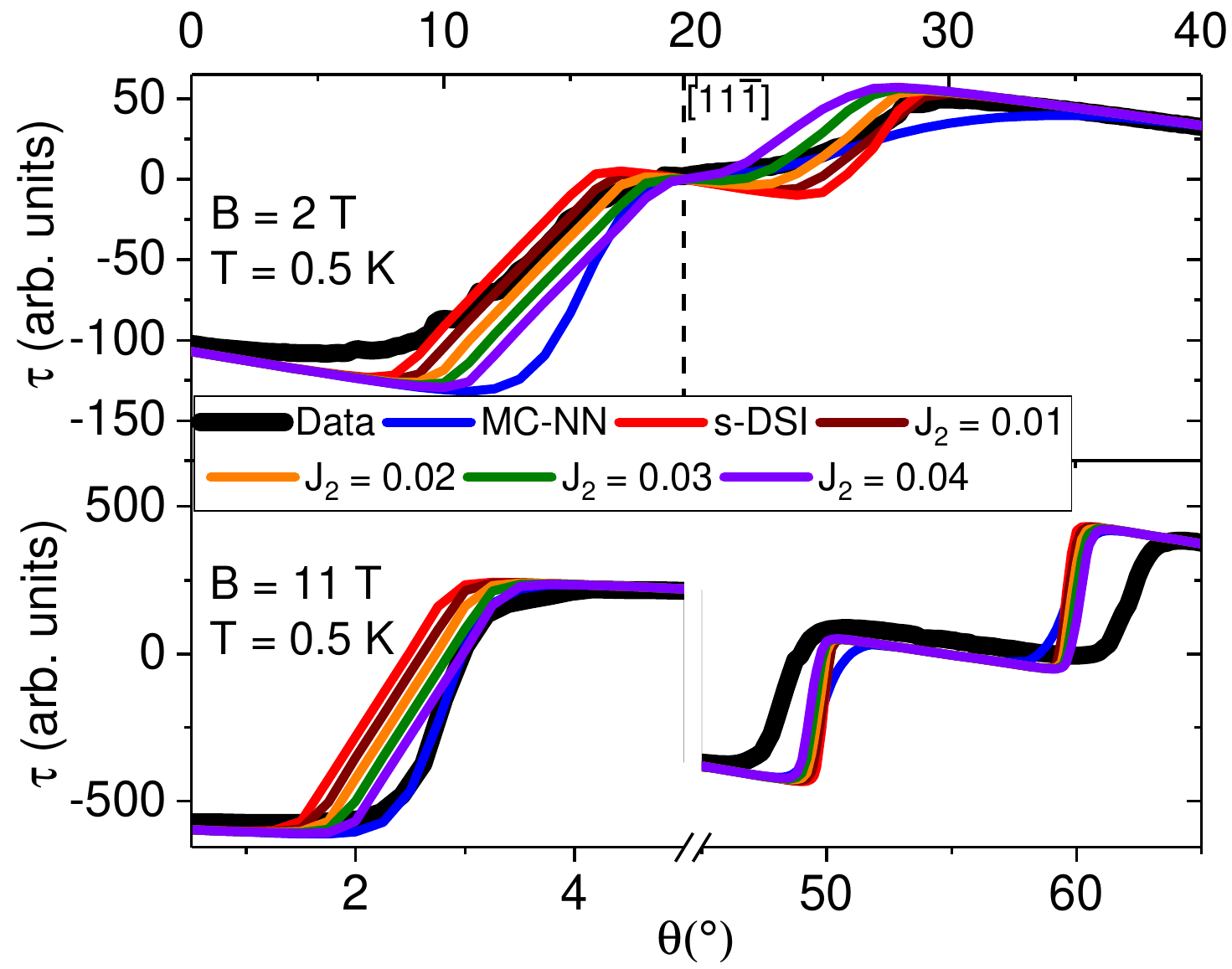}
    \caption{Simulated torque response as a function of angle, obtained from MC simulations using the nearest neighbor model (NN, blue curves), s-DSI model (red curves) and the generalized DSI model (s-DSI with J$_2$ = 0.01 meV, 0.02 meV, 0.03 meV, and 0.04 meV, wine, orange, green, and purple curves, respectively); measured CTM data (black curves) are taken at $T$ = 0.5~K in applied fields of 2~T (top panel) and 11~T (bottom panel). }
    \label{fig:MCJ2}
\end{figure}

In Supplementary Figure \ref{fig:MCJ2}, simulated torque curves are compared to the experimental curves taken at 0.5~K in 2 and 11~T, respectively. The results for a strictly nearest neighbor model (NN, blue curve) and for the s-DSI (long-range dipolar) model (including Ewald summation, loop moves, and demagnetization effects \cite{RogerGMelko}, red curve) are shown. The wine, orange, green, and purple curves were simulated with a more generalized DSI model using s-DSI model parameters and including various values for the $J_2$ interaction (0.01, 0.02, 0.03, and 0.04 meV). 

\section{Calculation of total energies of the spin-ice states and estimates of $J^{\alpha/\beta}_{eff}$}

We have calculated the total energy of each of the spin-ice states that we observed in the CTM experiment and the MC simulations. in the framework of a short-range model that includes $J_{1,eff}$, $J_{2,eff}$, $J_{3a,eff}$ and $J_{3b,eff}$. The zero field part of this effective Hamiltonian is, 
\begin{equation}
    H = J_{1,eff} \sum_{\langle i,j \rangle} \sigma_i \sigma_j + J_{2,eff} \sum_{\langle\langle i,j \rangle\rangle} \sigma_i \sigma_j + J_{3a,eff} \sum_{\langle\langle\langle i,j \rangle\rangle\rangle,a} \sigma_i \sigma_j +
    J_{3b,eff} \sum_{\langle\langle\langle i,j \rangle\rangle\rangle,b} \sigma_i \sigma_j 
    \label{eq:7}
\end{equation}
where $\langle \langle \langle i,j \rangle \rangle\rangle,a$ and  $\langle \langle \langle i,j \rangle \rangle\rangle,b$ refer to third nearest neighbors of $a$ and $b$ types respectively. (The $a$ type third nearest neighbors share a common neighbor and are thus connected to each other via a minimum of two ``hops'' on the pyrochlore lattice. The $b$ type third nearest neighbors do not share a nearest neighbor, they are instead related by a path involving a minimum of three hops on the pyrochlore lattice \cite{borzi,Henelius}.) 

We enumerated the first, second and third nearest neighbors (of both kinds $a,b$) for the pyrochlore lattice. (This was done numerically for a cubic unit cell of $16 \times 2^3 = 128$ sites). 
We computed the Zeeman energy and the interaction energy per 16-site unit cell for three phases: (2:2)$_0$ (with $\vec{m}||z$), (2:2)$_X$, and 3:1 ($+---$, i.e., 3-in 1-out). 

The sum of the interaction and Zeeman contributions (i.e., the second term in Supplementary Equation \ref{eq:NNHamlocalz} with $\vec{B}$ defined as ($h_x$,$h_y$,$h_z$)) gives us the total energy (which per 16-site cubic unit cell) is as below:
\begin{subequations}
\begin{eqnarray}
    (2:2)_0 &:&  -16 J_{1,eff} - 32 J_{2,eff} + 48 J_{3a,eff} + 48 J_{3b,eff} - \frac{16} {\sqrt{3}} g \mu_B h_z \\
     (2:2)_X &:&  -16 J_{1,eff} + 16 J_{3a,eff} + 16 J_{3b,eff} - \frac{8} {\sqrt{3}} g \mu_B (h_x+h_y) \\
      (3:1) &:&  +48 J_{3a,eff} + 48 J_{3b,eff} - \frac{8} {\sqrt{3}} g \mu_B (h_x+h_y+h_z) 
\end{eqnarray}
\end{subequations}
%The Zeeman energies written out as a function of the angle ($\theta$) between the applied magnetic field $\vec{B}$ and the magnetization $\vec{m}$ are
The magnetic field components as a function of the angle ($\theta$) are (see Supplementary Equations \ref{Bvectors}(a) and (b)):
\begin{subequations}
\begin{eqnarray}
    h_x &=& \Big(\frac{\cos \theta}{\sqrt{6}} +\frac{\sin \theta}{\sqrt{3}}\Big)B \\
     h_y &=& \Big(\frac{\cos \theta }{\sqrt{6}} +\frac{\sin \theta}{\sqrt{3}}\Big)B \\
      h_z &=& \Big(-\frac{2\cos \theta}{\sqrt{6}} +\frac{\sin \theta}{\sqrt{3}}\Big)B
\end{eqnarray}
\end{subequations}

\noindent To obtain the phase boundary of the $(2:2)_0 \Leftrightarrow (3:1)$ transition at zero temperature ($T=0$) we equate the total energies of the two phases (per 16-site unit cell). We get,
\begin{equation}
2 J_{1,eff} + 4 J_{2,eff} = \frac{1} {\sqrt{3}} g \mu_B (h_x+h_y-h_z)
\label{eq:q0_31}
\end{equation}
The phase transitions between symmetry related phases (2:2)$_0$ (with $\vec{m}||-z$) and 3:1 ($+++-$, i.e., 3-out 1-in) can be evaluated in a similar way. 
Note that the $J^a_{3,eff}$ and $J^b_{3,eff}$ terms cancel out for this transition. Thus, $J_3$ interactions do not affect this transition. 

  %HJC edited this: Using a similar short range model compared to what was considered in 
   The effective parameters can be expressed in terms of the original generalized DSI model (now incorporating second and third nearest neighbors with couplings $J_2$, $J_{3a}$, and $J_{3b}$) following Ref.~\cite{borzi}, who considered the case of $J_2=J_{3a}=J_{3b}=0$, but with the opposite sign convention of $J_1$ compared to our work.%, we can extract estimates for the value of $J^{\alpha/\beta}_{eff}$ seen in the CTM experiment. 
 The relations are,
\begin{subequations}
\begin{eqnarray}
    J_{1,eff} &\equiv& \frac{J_1 + 5D}{3} \\
    J_{2,eff} &\equiv& \frac{J_2 - D/\sqrt{3}}{3}\\
    J_{3a,eff} &\equiv&  -J_{3a} - \frac{D}{8} \\
    J_{3b,eff} &\equiv&  -J_{3b} + \frac{D}{8}   
\end{eqnarray}\label{J2eff}
\end{subequations}
The factor of $1/3$ arises from the geometric factor $z_i \cdot z_j = -1/3$ for two spins on different sublattices as is the case for nearest and second nearest neighbors. The third nearest neighbors are on the same sublattice for which $z_i \cdot z_i =1$. Note that this model truncates the dipolar interaction, and is thus expected to be only qualitatively accurate.  
Based on Supplementary Equations \ref{J2eff} and \ref{eq:q0_31} we can extract a $J^{\alpha}_{eff}$ that would correspond to this combination of interaction terms.% We ask: what effective nearest neighbor "$J_{eff}$" would mimic the effects of this combination of $J's$ in our effective model with further-neighbor couplings. We obtain, 
\begin{equation}
    J^{\alpha}_{eff} \equiv J_{1,eff} + 2J_{2,eff} = \frac{1}{3} \Big( J_1 + 2 J_2 + \Big(5- \frac{2}{\sqrt{3}} \Big) D \Big)
    \label{eq:12}
\end{equation}
For $J_1 =-0.13499$ meV (-1.56~K), $D=0.121567$ meV (1.41~K), and $J_2 = 0.030 $ meV (0.35~K) we get $J^{\alpha}_{eff}$ = 1.52~K, which is lower than $J_{1,eff}$=1.83~K, encouragingly it is in the same ballpark as that measured in the CTM experiment. Hence, once the interaction terms are known, we can rearrange Supplementary Equation~\ref{eq:q0_31} and plot $B$ as a function of $\theta$, which represents the phase boundary at which the transition occurs (see Supplementary Figure \ref{fig:boundary} (a)). We plot the phase boundary for three combinations of $J_1$, $J_2$, and $D$.     
\\

\noindent We repeat the analysis for the (2:2)$_X \Leftrightarrow$ (3:1) transition. 
On equating their energies, we get
\begin{equation}
2 J_{1,eff} + 4 J_{3a,eff} + 4 J_{3b,eff} = \frac{1} {\sqrt{3}} g \mu_B h_z
\label{eq:qx_31}
\end{equation}
Based on Supplementary Equations \ref{eq:qx_31} and \ref{J2eff}, we can extract a $J^{\beta}_{eff}$ that would correspond to this combination of interaction terms.
\begin{equation}
    J^{\beta}_{eff} \equiv J_{1,eff} + 2 (J_{3a,eff}+ J_{3b,eff}) = \frac{J_1 + 5 D}{3} -2 (J_{3a} + J_{3b}) 
    \label{eq:14}
\end{equation}
We can thus use the field dependent critical angles obtained by CTM to constrain $J_{3a} + J_{3b} \approx -0.014$ meV ($-0.16$~K) by solving Supplementary Equation~\ref{eq:qx_31} for multiple field strengths and looking at the value that best agrees with the experimental data. This value of $J_{3a} + J_{3b}$ corresponds to a $J^{\beta}_{eff}$  of about 2.15~K. Note however that this argument alone does not constrain $J_{3a}$ and $J_{3b}$ individually. In Supplementary Figure\ref{fig:boundary} (b), the calculated phase boundaries are shown for the (2:2)$_X \Leftrightarrow $(3:1) transition plotted along side the critical angles extracted for the CTM data. %HJC commented this: boundaries are shown for a %NN-model and for a model that includes %($J_{3a}$+$J_{3b}$)=-0.16~K. 

\begin{figure}
    \centering
    \includegraphics[width=1.0\textwidth]{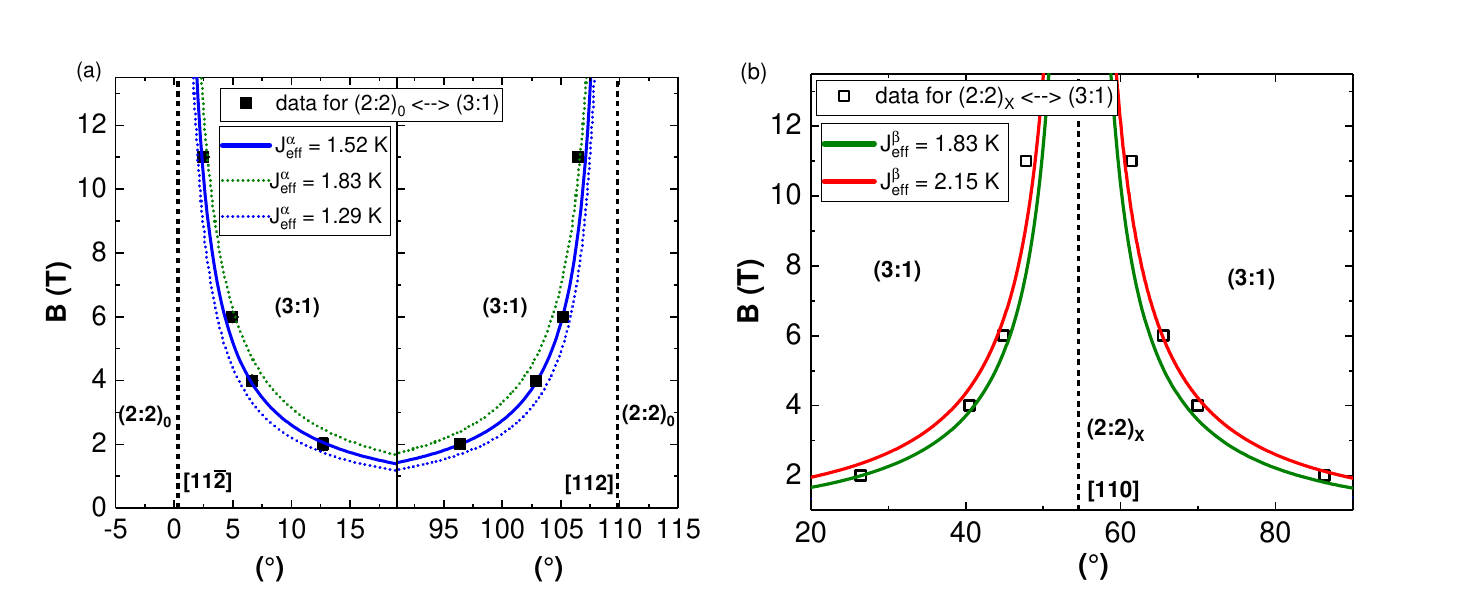}
    \caption{(a) Critical angles extracted from CTM data as a function of applied field for the \hkl(1-10) plane rotation for the (2:2)$_0$ $\Leftrightarrow$ (3:1) transitions (solid squares). This is the same data as was plotted in the main text Figure 4(a). Calculated phase boundaries for this transition using Supplementary Equations \ref{eq:q0_31} and \ref{eq:12}, the solid blue line represents the model using $J^{\alpha}_{eff}$ = 1.52~K (with $J_1$=-1.56~K, $J_2$ = 0.35~K, and $D$=1.41~K); the green dotted line represents the model for which $J_{2,eff}$=0 and $J_{1,eff}$=1.83~K (with $J_1$=-1.56~K and $D$=1.41~K (the NN-model)); the blue dotted line represents the model using Supplementary Equation \ref{eq:12} but (with $J_1$=-1.56~K, $J_2$=0,  and $D$=1.41~K, i.e., dipolar interactions are truncated at the second nearest neighbor). (b)  Critical angles extracted from CTM data as a function of applied field for the \hkl(1-10) plane rotation for the (2:2)$_X$ $\Leftrightarrow$ (3:1) transitions (open squares). This is the same data as was plotted in the main text Figure 4(a). Calculated phase boundaries for this transition using Supplementary Equations \ref{eq:qx_31} and \ref{eq:14}. The solid green line represents the model using $J^{\beta}_{eff}$ = 1.83~K with $J_1$=-1.56~K, ($J_{3a}$+$J_{3b}$)=0,  and $D$=1.41~K (the NN-model). The red solid line represents the model using Supplementary Equation \ref{eq:14} with $J_1$=-1.56~K, $D$=1.41~K, and ($J_{3a}$+$J_{3b}$)=-0.16~K.   }
    \label{fig:calcE}
     \label{fig:boundary}
\end{figure}

\section{Single crystal X-ray diffraction} 
Single crystal X-ray diffraction was used to fully characterize the structure of the HTO single crystals. A small specimen was cut directly from a larger piece used for torque measurements and mounted to the end of a glass fiber using clear adhesive. A full quadrant of X-ray diffraction data were collected under ambient conditions using $\omega$-scans with 1\degree frame widths to a resolution of 0.4 \AA, equivalent to 2$\theta \sim$ 125 \degree, on an Oxford-Diffraction Xcalibur-2 CCD diffractometer equipped with a graphite-monochromated MoK$_{\alpha}$ source. Reflections were recorded, indexed and corrected for absorption using the Oxford-Diffraction CrysAlisPro software \cite{Crysal}, initial structure determination was performed using SHELXS \cite{SHELX} and final cycles of structure refinement were carried out in CRYSTALS \cite{CRYSTAL}. Model refinement was performed against F$^2$ (i.e., structure factor intensities), and high-quality data allowed unconstrained refinement of all atom positions and anisotropic displacement ellipsoids.

Final structural refinement results and structure factors  have been deposited with the joint CCDC/FIZ Karlsruhe online deposition service;  these files are available for download under deposition nr. CSD-2172269 \cite{CSD}.

% \begin{table}[htpb]
%\begin{center}
%\begin{tabular}{|c|c|c|c|c|c|}
%\hline
%Model                  &  $J_1$ (K) &D(K)& J$_{eff}$  (K) & $J_2$ (K)&  $J_{3a}$ ($J_{3b}$) (K) \tabularnewline
%\hline
%NN  &   5.40             & 0         &   1.80 &     	&			    \tabularnewline
%s-DSI  &   -1.56             &1.41         &   1.83 &     	&			    \tabularnewline
 %s-DSI  &   -1.56            &  1.41        &  1.61 &    & 				    \tabularnewline
%g-DSI &   -1.56  & 1.41       &  1.61  &  ?   		&  ?	    %\tabularnewline
%\hline
%\end{tabular}
%\end{center}
%\end{table}
\renewcommand\refname{Supplementary References} 
\bibliography{TorqueHTO}